\documentclass{article}
\usepackage{arxiv}

\usepackage[utf8]{inputenc} 
\usepackage[T1]{fontenc}    
\usepackage{hyperref}       
\usepackage{url}            
\usepackage{booktabs}       
\usepackage{amsfonts}       
\usepackage{nicefrac}       
\usepackage{microtype}      
\usepackage{lipsum}
\usepackage{subcaption}
\usepackage{graphicx}
\usepackage{natbib}
\usepackage{booktabs}
\usepackage{enumitem}
\usepackage{mathtools}

\title{Bridge Modal Identification using Acceleration\\ Measurements within Moving Vehicles}

\author{
  Soheil S. Eshkevari\thanks{ses516@lehigh.edu} \\
  Department of Civil and Environmental Engineering\\
  Lehigh University\\
  Bethlehem, PA 18015 \\
  \texttt{ses516@lehigh.edu} \\
   \And
  Thomas J. Matarazzo\thanks{tomjmat@mit.edu} \\
  Senseable Lab\\
  Massachusetts Institute of Technology\\
  Cambridge, MA 02139 \\
  \texttt{tomjmat@mit.edu} \\
  \And
 Shamim N. Pakzad \\
  Department of Civil and Environmental Engineering\\
  Lehigh University\\
  Bethlehem, PA 18015 \\
  \texttt{pakzad@lehigh.edu} \\
}

\begin{document}
\maketitle
\begin{abstract}
Vehicles commuting over bridge structures respond dynamically to the bridge's vibrations. An acceleration signal collected within a moving vehicle contains a trace of the bridge's structural response, but also includes other sources such as the vehicle suspension system and surface roughness-induced vibrations. This paper introduces two general methods for the bridge system identification using data exclusively collected by a network of moving vehicles. The contributions of the vehicle suspension system are removed by deconvolving the vehicle response in frequency domain. The first approach utilizes the vehicle transfer function, and the second uses ensemble empirical modal decomposition (EEMD). EEMD is a geometric based blind source separation (BSS) tool that is able to extract sources out of a statically or convolutionally mixed signal. Next, roughness-induced vibrations are extracted through a novel application of second-order blind identification (SOBI) method. After these two processes the resulting signal is equivalent to the readings of mobile sensors that scan the bridge's dynamic response. Structural modal identification using mobile sensor data has been recently made possible with the extended structural modal identification using expectation maximization (STRIDEX) algorithm. The processed mobile sensor data is analyzed using STRIDEX to identify the modal properties of the bridge. The performance of the methods are validated on numerical case studies of a long single-span bridge with a network of moving vehicles collecting data while in motion. The analyses consider three road surface roughness patterns. Results show that the proposed algorithms are successful in extracting pure bridge vibrations, and produce accurate and comprehensive modal properties of the bridge. The study shows that the proposed transfer function method can efficiently deconvolve the linear dynamics of a moving vehicle. EEMD method is able to extract vehicle dynamic response without a-priori information about the vehicle. In addition, proposed identification methods provide secondary information about the roughness pattern and the vehicle. This study is the first proposed methodology for complete bridge modal identification, including operational natural frequencies, mode shapes and damping ratios using \textit{moving vehicles sensor data}. 

\end{abstract}


\section{Introduction}
\label{S:0}

Advancements in sensor technology and data acquisition techniques have played a major role in bringing the civil engineering community towards more frequent and more accurate condition assessments of structures. Structural health monitoring (SHM) methods have been quick to evolve with technologies; most notably, wireless sensor networks became attractive alternatives to wired ones \cite{lynch2006summary,ni2009technology,pakzad2008design}. In terms of system identification techniques, many statistical frameworks have been proposed and verified \cite{andersen1999comparison,juang1985eigensystem,gul2009statistical}. Yet, the measurement process in SHM still adheres to the fixed sensor paradigm. In this framework, sensors are installed at certain locations on the structure, thus the spatial information in the data is restricted to these particular points. Ultimately, this measurement approach can limit researchers' ability to understand structural condition and performance. For example, in system identification (SID), it is known that the spatial resolution of the identified mode shapes is directly impacted by the number of sensing nodes and their arrangement \cite{matarazzo2016structural}. \par

SID applications that target higher resolution mode shapes have used dense arrays of fixed sensors networks \cite{pakzad2008design, dorvash2014application,zhu2012wireless}; while such networks are able to provide an improvement in spatial information, the equipment, setup, and maintenance costs associated with dense sensor networks effectively make this type of information inaccessible. In addition, complex networks have more complicated communication and processing tasks since more data need to be transmitted and analyzed, requiring more advanced communication technologies. Recently, real-time monitoring has been studied using internet of things (IoT) for data transmission and storage clouding \cite{smarsly2011multiagent,kijewski2012smartsync,zhang2016senstore}. These technologies have eased the application of complex sensing networks, while the spatial resolution problem is not addressed yet.\par

The cost inefficiency of highly dense sensor networks has motivated researchers to determine more optimal sensor layouts, e.g., those that minimize network complexity while attaining a level of information that is appropriate for the application. This approach has been taken for both damage detection \cite{guo2004optimal,kim2000structural} and structural modal identification \cite{chang2015optimal,meo2005optimal} purposes. While important and highly practical, this strategy does not resolve the scalability problems that are inherent to fixed sensor networks because of two concerns: requiring more sensors for more spatial information and requiring dedicated sensor networks for each structure to be assessed. \par

Mobile sensor networks offer numerous advantages compared to the conventional stationary sensing scenario. Overall, mobile sensors have low setup costs, collect spatial information efficiently, and no dedicated sensors to any particular structure. Most importantly, mobile sensors can capture comprehensive spatial information using few sensors. \par

\subsection{Toward infrastructure vibration crowdsourcing} 

The advantages of mobile sensing combined with the ubiquity of smartphones with IoT connectivity have motivated researchers to think of automobiles as large-scale sensor networks. Recently, studies were conducted to show the suitability of mobile sensors and smartphones for environmental assessment in urban areas. Recent researches \cite{eriksson2008pothole,alessandroni2014smartroadsense,kumar2016community,anjomshoaa2018city}
have incorporated smartphone vibration data for road pothole detection and road surface condition assessment purposes, and successfully examined their platform experimentally. \citet{feng2015citizen} studied substitution of stationary sensors with smartphones for identifying dynamic characteristics of structures. However, the sensing network consisted of fixed nodes. \citet{matarazzo2018crowdsensing} studied crowdsensing possibilities created by smartphone abundance. In this study, smartphones were used for data collection while driving over a bridge and by frequency-domain analysis, bridge natural frequencies were identified. They showed that smartphones are viable moving sensors and easy to use; however, the result was limited to frequency informations and was not a comprehensive SID. The paper emphasized on the huge information potential offered by ubiquitous smartphones in moving vehicles.\par

Despite the promises of scalability, there remains a need to develop practical analytical procedures for a comprehensive SID that are applicable to mobile sensor and smartphone data. The recent studies considering mobile sensor networks have mostly demonstrated partial modal identification(either frequency, damping, or mode shapes). At this time, STRIDEX by \citet{matarazzo2018scalable} is the one of the procedure that is capable of performing a complete identification using mobile sensor data. \citet{sadeghi2020EM} has recently proposed an alternative approach as well with a different sensing setting. \par 

\subsection{Review of system identification using mobile sensor networks}

Mobile sensor networks offer more scalable and flexible data acquisition when compared to fixed sensor networks; yet the resulting data sets are fundamentally different and require special consideration \cite{matarazzo2016truncated}. Mobile sensing data fall under the dynamic sensor network (DSN) data classification and can be mapped exactly to the equation of motion through the truncated physical state-space model (TPM). In a TPM, the fundamental assumption is that the measurements at physical locations are truncated results of a coordinate transform from modal coordinates at every time sample. Equation \ref{eq:StateSpace} shows the state-space model of a dynamic structure: \par

\begin{equation}\label{eq:StateSpace}
\begin{split}
x_k = Ax_{k-1}+\eta_k \\
y_k = \Omega_k Cx_k+\nu_k
\end{split}
\end{equation}

In this equation, $x_k$ (with $x_1\sim N(\hat{\mu},\hat{V})$) is the state vector containing the structural response at all degrees of freedom (DOFs), $y_k$ is the observation vector which includes the structural responses at a subset of all DOFs. $A\in \mathbb{R}^{pN_\alpha\times pN_\alpha}$ and $C\in \mathbb{R}^{N_0\times pN_\alpha}$ are state and observation matrices, respectively. $N_\alpha$ is the number of virtual probing nodes and $N_0$ is the size of observation vectors. $p$ is a user-defined model order, expanding number of observation channels to a desired number of states. $\eta_k\sim N(0,Q)$ and $\nu_k\sim N(0,R)$ are systematic and sensing noises, which are commonly modeled as uncorrelated Gaussian white noise with covariance matrices $Q$ and $R$. Finally, $\Omega_k$ is called mode shape regression (MSR) function, and essentially is a known 3D array that for each time step, maps moving sensors data to virtual probing data by an approximation. Further mathematical details for MSR function can be found in \cite{matarazzo2016truncated}. In addition, a more advanced signal reconstruction method is proposed by \citet{eshkevari2020signal} which enables data estimation on probing locations under high irregularities in the mobile sensors network. \par

The governing state-space model (Equation \ref{eq:StateSpace}) of the system is time variant with six time-invariant parameters $\{A, C, \hat{\mu},\hat{V}, Q, R\}$ and one time variant parameter $\{\Omega_k \}$, which can be approximated using a \textit{sinc} basis function. The structural identification using expectation maximization (STRIDEX) method was proposed to identify the parameters of the TPM and extract modal properties of the structural system, which include high-resolution mode shapes. The accuracy and performance of STRIDEX was validated using both synthetic and experimental data \cite{matarazzo2018scalable}. Most notably, it was shown that two mobile sensors can produce a mode shape estimation with over 240 points.\par




The STRIDEX \cite{matarazzo2018scalable} was developed to determine the maximum likelihood estimates (MLE) of the TPM (Equation \ref{eq:StateSpace}). This is achieved using the expectation maximization, a method which computes iteratively the conditional expectation of the unobserved state variable and its covariance matrices. In STRIDEX, the time invariant parameters are combined into one \textit{super-parameter}. In the expectation (E) step, given the observed measurement data, state vectors are estimated using Kalman filtering \cite{ristic2004beyond} and Rauch-Tung-Striebel (RTS) smoothing \cite{sarkka2008unscented}. Then the conditional expectation of the log-likelihood function of TPM is maximized to yield an updated super-parameter estimate (maximization (M) step). This procedure continues until log-likelihood function value changes less than a predefined threshold. Detailed explanations and mathematical proofs are available in \cite{matarazzo2018scalable}. \par

While successful, in previous applications of STRIDEX, it was assumed that the mobility mechanism did not contaminate the measurement process. In other words, considering the case of a sensor within a moving vehicle, the vehicle dynamics and road profile effects were not considered. In practice, a sensor within a moving carrier cannot capture the pure dynamic response of the bridge. These are a mixture from several sources, primarily roughness-induced vibrations and vehicle suspension vibrations. The following section presents an overview of relevant studies considered real-world data subjected to the vehicle-bridge interaction problem.

\subsection{Bridge system identification using dynamic sensor network data}

Realistically, moving sensors are placed into a carrier, which itself is a mechanical system. Vehicle-carried sensors in fact are collecting responses of a vehicle suspension system under bridge vibrations entering via tires. Researchers have been studying vehilce-bridge interaction, mostly assuming one passing vehicle as the loading of the bridge \cite{cantero2019experimental,chang2014variability}. Indirect bridge monitoring considers the extraction of bridge properties, in a general sense, from the response of a moving vehicle; however, the specific objectives have varied throughout these studies. Some have aimed to obtain partial modal information, e.g., frequencies only \cite{yang2004extracting, lin2005use, yang2009extracting, siringoringo2012estimating}, or mode shapes \cite{malekjafarian2014ident}, or damping ratio of the fundamental mode \cite{gonzalez2012identification} given a special bridge excitation. Others have targeted attributes that are related to modal properties such as mode shape squares \cite{zhang2012damage} and stiffness indicators \cite{malekjafarian2017use}. Very limited studied have been done on bridge identification using mobile data under ambient loading \cite{marulanda2016modal}, with some constraints. \par
A frequency-domain bridge identification based on mobile sensor data was introduced by \citet{yang2004extracting}, which used the acceleration time history of a vehicle crossing a bridge to determine its fundamental frequency. A parametric study was done to capture effects of vehicle speed and bridge mechanical properties and numerically verified by finite element analysis. The bridge loading was moving point loads applied by the data collector vehicle. The research was experimentally validated by \citet{lin2005use}, in which a moving setup, consisting of a tractor and trailers passing with various speeds over a bridge and further investigations were conducted by \citet{yang2009extracting}. A comprehensive finite element study was performed by \citet{siringoringo2012estimating}, in which a detailed bridge-vehicle interaction model (VBI) was built and vehicle vibrations caused by bridge dynamic motions were used for frequency-domain SID. The study also was backed up by an experiment on a bridge in Japan. However, the methodology was only able to capture the first natural frequency and limited to a given moving load scenario. \par
The same framework (VBI) was studied by \citet{gonzalez2012identification} in order to estimate bridge damping ratios using moving acceleration records, assuming Rayleigh damping. The bridge was loaded by a passing sensor carrier. The study suggests an optimization procedure to tune the most fit damping ratio for the first mode of a bridge based on its geometry and mechanical properties. A SID procedure to refine identified mode shapes using time-frequency signal representation between fixed and mobile sensors was proposed by \citet{marulanda2016modal}. The algorithm showed high resolution mode shapes extraction under numerical noise-free and experimental data. However, mobile sensing data were assumed to be pure structural vibrations in both numerical and experimental cases. This study required stationary nodes, however, it was assuming a more general loading scenario. \par 

\citet{zhang2012damage} proposed a method to extract mode shapes of a plate and beam structures from acceleration response of a traveling device with a tapping instrument as a loading apparatus. In this study, a single sensor was used and good mode shape resolution could be achieved. The mode shapes are found by exciting the bridge by frequencies close to known natural frequencies. In addition, the method does not assume any contamination caused by other sources. \par

As a more realistic case, \citet{malekjafarian2014ident} used short time frequency domain decomposition for estimating bridge mode shapes using vehicle-carried sensing data. This study also considered the bridge under a special moving load resulted by a passing vehicle. The method showed promising results in the case of a very smooth roadway (no roughness considered). For the second phase, to eliminate roughness-caused vibrations, the residual of axle signals from two successive trailers was used as the input. The approach was successful for identifying mode shapes of first two modes. However, the spatial resolution was low and vehicle properties were not fully given. Alternatively, the same researchers extended the idea by adding a tapping device exciting the bridge with the frequency close to the bridge natural frequencies \cite{malekjafarian2017use}. For diminishing road roughness effect, the same technique of subtracting measurements of two axles was used and higher resolution mode shapes were identified. The method needs specialized vehicles that are pulling auxiliary parts, which are not common. In addition, the study aimed to identify mode shapes, not a complete modal identification. Note that neither of these studies considered ambient bridge excitation. \par

\subsection{Problem definition}

Bridge vibrations are transmitted into the vehicle cabin as it travels across the bridge. The goal is to extract bridge dynamic information from vehicle accelerations; however, these measurements are subject to the vehicle-bridge interaction problem, i.e., polluted by undesired signals. Mathematically, the TPM developed for the mobile sensing problem shown in Equation \ref{eq:StateSpace} can be formulated to consider this particular measurement process as shown in Equation \ref{eq:prbdef}:

\begin{equation}\label{eq:prbdef}
\begin{split}
x_k = Ax_{k-1}+\eta_k \\
y_k = y_k^{br} = \Omega_k Cx_k+\nu_k \\
y_k^{vbi} = y^k_{br} + y_k^{if} \\
y_k^{act} = f(y_k^{vbi}+y_k^{rgh},y_k^{eng})
\end{split}
\end{equation}

where $y_k = y_k^{br}$ and $y_k^{if}$ are the pure bridge response and bridge response under vehicle-bridge interacting force, respectively. The algebraic sum of these two components constitute $y_k^{vbi}$, when the bridge interacts with the vehicle. The bridge-vehicle interaction in long span bridges can be modeled in detail as proposed in \cite{zhou2015fully,camara2019complete}. In this study since a bridge under ambient load is considered, the vehicle-bridge interaction part $y_k^{if}$ is significantly smaller that $y_k^{br}$, resulting $y_k^{vbi} \simeq y_k^{br}$. This approximation has been demonstrated by comparing coupled and uncoupled responses of the bridge and the vehicle in the preliminary stage of the analysis. The mobile sensor measurement $y_k^{act}$ is a complex mixture of multiple simultaneous phenomena which includes the idealized mobile sensor data $y_k^{br}$. $y_k^{rgh}$ is the profile roughness displacements at time $k$. $y_k^{eng}$ is the engine-induced vibration, which presents in the vehicle response, is often actively controlled and not sensible \cite{elliott1990active}, thus is neglected in this study. Most importantly, the final response $y_k^{act}$ is a nonlinear function of the linear mixture of these three components. The function $f(\cdot)$ is the \textit{convolution} of the vehicle impulse response by the input signal.

\begin{figure}[!htbp]
\centering\includegraphics[width=0.8\linewidth]{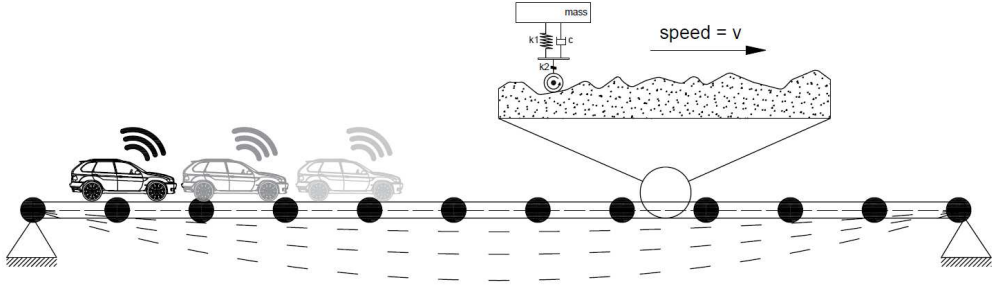}
\caption{Schematic illustration of vehicle-carried sensing}
\label{prb_def}
\end{figure}

The objective is to extract $y_k^{br}$ from the actual vehicle response $y_k^{act}$ and implement STRIDEX to provide a complete bridge SID.

\subsection{Motivation}

 Given the state of the literature, a method that can simultaneously process data from multiple moving sensors within vehicles makes an important contribution to crowd-based systems. We found that the significant challenge in large-scale “crowdsensing” campaigns is a methodological gap: the lack of an “end-to-end” system identification process that accounts for unwanted vehicle vibrations. Therefore the goal of this paper is to present a set of methodologies that decompose the measurements recorded within the vehicle cabin into individual components. Blind source separation (BSS) as a signal processing tool provides methodologies to decompose statically or convolutionally mixed signals to original sources without any knowledge about them in advance \cite{cardoso1998blind}. The BSS problem is ill-conditioned, since both sources and mixing matrix are unknown; however, many assumptions have made to constrain the problem. \par

Recently, \citet{sadeghi2017imac} showed that for a vehicle with a rigid suspension system, second-order blind identification (SOBI) is capable of separating roughness-induced vibration and bridge dynamics. That is, the bridge vibrations were extracted from a pair of vehicle measurements collected over a certain portion of the bridge length. However, the study did not consider realistic vehicle dynamics. This study incorporates a vehicle dynamics model and the technical goal is to remove both vehicle and roughness effects from raw acceleration measurements using signal deconvolution and blind source separation (BSS) techniques, e.g., EEMD \cite{huang1998empirical} and SOBI \cite{belouchrani1997blind}. The objective of this phase is to decompose the highly contaminated raw signal collected by the vehicle sensor and extract bridge vibrations. Next, STRIDEX is implemented to process the mobile sensing data extracted and determine the structural modal properties (frequencies, damping ratios, and mode shapes). \par

\begin{figure}[!ht]
\centering\includegraphics[width=1.0\linewidth]{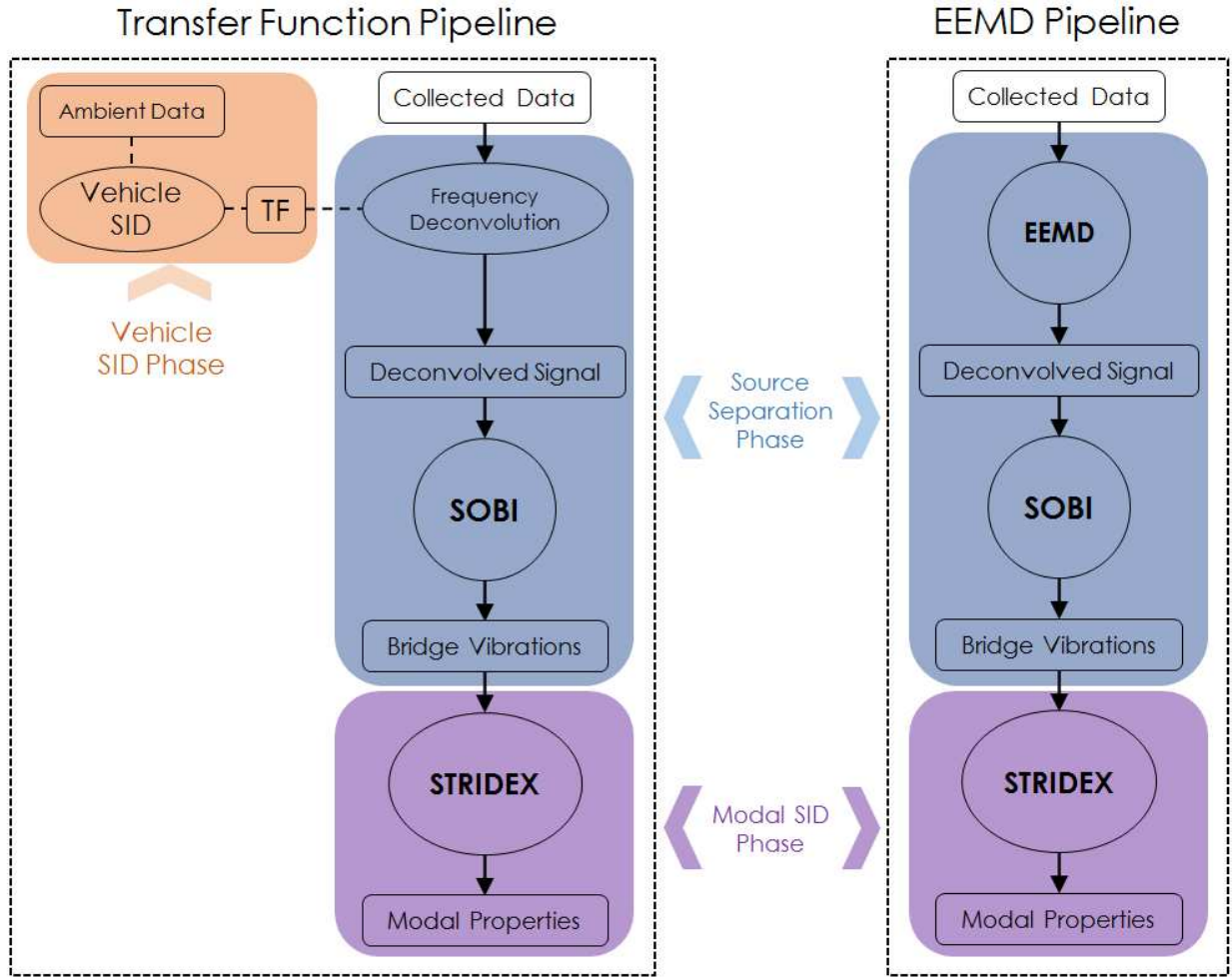}
\caption{Flowcharts for two proposed pipelines}
\label{methods_fc}
\end{figure}

In this study, two approaches are proposed to extract the bridge response from vehicle measurements (as shown in Figure \ref{methods_fc}): the Transfer function (TF) approach and the EEMD approach. Each procedure consists of two phases: deconvolution of the vehicle dynamical effect and removal of roughness-induced vibrations. The key difference between the two approaches is the method by which the vehicle vibrations are separated. The TF approach uses the vehicle's frequency response function (FRF) for the deconvolution, while the EEMD approach estimates the sources of a signal by using trend extraction. \par

In either case, after deconvolution, the nonlinear function $f(\cdot)$ of Equation \ref{eq:prbdef} is inverted, so that the remainder is the argument of the function i.e. a linear mixture of the bridge dynamic vibrations $y_k^{br}$ and roughness profile $y_k^{rgh}$, neglecting engine-induced noises. Therefore, the signal still needs to be separated to extract $y_k^{br}$. This objective is being done by applying SOBI which is a robust solution for un-mixing linearly superposed sources. \par

In the following sections, these procedures are detailed and their performances are evaluated using realistic numerical simulations. The main contributions of this paper are as follows:

\begin{enumerate}
	\item The proposed approaches are the first that provide a comprehensive modal identification of a bridge using acceleration measurements from moving vehicles.
	\item The approaches are proposed to be used for bridges under ambient loading, with minimal assumptions on bridge type and vehicle characteristics.
	\item These approaches are robust for severely rough road profiles as long as the vehicle acts linearly within the frequency band considered.
\end{enumerate}

In Section \ref{sec:vcl_sim}, a simulation that has been used for numerical assessment is demonstrated and methods processes are presented. Later in Section \ref{sec:brg_ext}, the procedures for signal decontamination are performed step by step and results are presented. In the next section, the purified signals are fed into the SID algorithm, STRIDEX, and results are discussed. In Section \ref{comm_car} the vehicle properties are changed from a vehicle with customized mechanical properties that is suitable for sensing (in terms of having disjoint frequency contents with the bridge) to a vehicle with regular mechanical properties and results are presented. Finally, in Section \ref{sec:TF_est} a more practical version of the first method (deconvolution using vehicle FRF) is proposed and evaluated. 

\section{Approaches to Extract the Bridge Response from Measurements within Moving Vehicles}\label{sec:Deconv}

This section presents two general approaches to extract the bridge acceleration response from measurements within a moving vehicle. The primary differences in the approaches lie within the deconvolution process. Both approaches implement second-order blind identification (SOBI) after signal deconvolution. In addition to the system transfer function approach, a modified version of that is presented in which the deconvolution phase is eased significantly.

\subsection{Method 1: Deconvolution using the system transfer function}

\begin{figure}[!htbp]
\centering\includegraphics[width=0.4\linewidth]{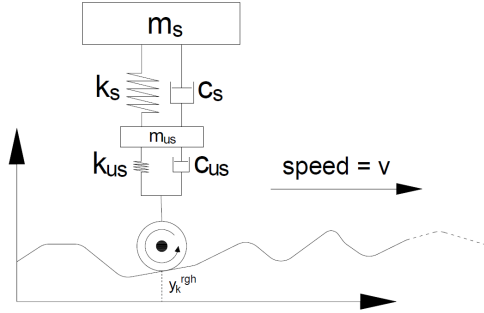}
\caption{Quarter-car model for vehicle suspension system}
\label{qrt_car}
\end{figure}

To remove vehicle dynamics effect from the collected response, the direct approach is to apply deconvolution in frequency domain, for which an accurate description of the vehicle suspension system is needed. In cases where this information is  not readily available a vehicle SID must be performed. The vehicle suspension system can be modeled efficiently as a linear quarter-car system, shown in Figure \ref{qrt_car} \cite{cebon1999handbook}.

where $m_s$ and $m_{us}$ are sprung and unsprung mass of a quarter-car. $c_s$, $c_{us}$ and $k_s$, $k_{us}$ are corresponding damping and stiffness for masses, respectively. At each time step, $y_k^{rgh}$ is the displacement input of the system caused by the road profile roughness and vehicle response is collected while attached to the $m_s$. For a regular road, $y_k^{rgh}$ is the input of the vehicle system, while for a vehicle-bridge interacting model, the vehicle input also contains the bridge vibrations $y_k^{brg}$. The model is a two DOF dynamic problem and a full description of the system can be expressed by a two DOF state space equation set.

Equations \ref{eq:1} and \ref{eq:2} show the governing dynamic equations of a vehicle with respect to the bridge vibrations and road profile roughness. In Equation \ref{eq:1}, $h_k$ is the quarter-car model impulse response in time domain, which in frequency representation is equivalent to the system transfer function (TF). Given system model and responses at both DOFs, the system can be identified using an output-only SID algorithm. After determining the vehicle's modal properties, the TF can be generated in frequency domain ($H(\omega)$) and then using Equation \ref{eq:3}, the vehicle input can be estimated. This input is yet to be processed for extracting $y_k^{brg}$, however, it is now a linear mixture of two components shown in Equation \ref{eq:2}. Note that the TF here is equivalent to the frequency response function (FRF) in the structural dynamics control literature. 

\begin{equation}\label{eq:1}
y_k^{obs}=h_k*(y_k^{brg}+y_k^{rgh})
\end{equation}
\begin{equation}\label{eq:2}
y_k^{inp}=y_k^{brg}+y_k^{rgh}
\end{equation}
\begin{equation}\label{eq:3}
Y^{obs}(\omega)=H(\omega)\times Y^{inp}(\omega)
\end{equation}

\subsection{Deconvolution using the approximated system transfer function}

In Method 1, a complete description of the vehicle system is possible only when the system response at all DOFs are collected. This means that in order to construct the identified TF, measurements collect vehicle response not only inside the vehicle (sprung channel), but on the unsprung mass level (tire channel), which is a hard task. An alternative approach is to assuming vehicle suspension mode shapes as given, collect data only inside the vehicle. In this scenario, the sprung channel is adequate to generate vehicle TF, since output-only SID tools can identify natural frequencies and damping ratios using only one data channel. This approach is practically more desirable and will be analyzed in Section \ref{sec:TF_est}. 

\subsection{Method 2: Deconvolution using ensemble empirical modal decomposition}\label{sec:EEMD}

Separation of mixed signals to sources is the primary focus in the field of blind source separation (BSS). BSS is a signal processing topic that has been widely studied for both static and convolved mixtures \cite{cardoso1998blind}. The process of BSS was used for structural dynamics application in \cite{poncelet2007output} and \cite{kerschen2007physical} and theoretical equivalence of the application to the classical signal un-mixing problem was explained. Empirical modal decomposition (EMD) \cite{huang1998empirical} is a single-channel source separation technique that ideally is able to extract convoluted mixtures, as well as linear mixtures. The method is a geometrical based process for single-channel source separation which in some special circumstances, is successfully able to separate even non-stationary or nonlinearly mixed sources. Components extracted from EMD out of the signal represent embedded oscillatory trends in the original signal. EMD in its initial form suffers from frequency leakage and aliasing between components  \cite{huang1998empirical}. As an extension, an ensemble of EMDs (EEMD) is proposed and is more common for blind source separation applications. EEMD introduces additional noise to the signal and perform EMD procedure, and repeat these steps multiple times. By averaging EMD results, final components are found. In this study, EEMD is being used for extracting the vehicle dynamic response out of the mixture. Methods 1 and 2 will be performed and discussed further in Section \ref{sec:brg_ext}.

\subsection{Source separation on the linear mixture}

\begin{figure}[!htbp]
\centering\includegraphics[width=0.4\linewidth]{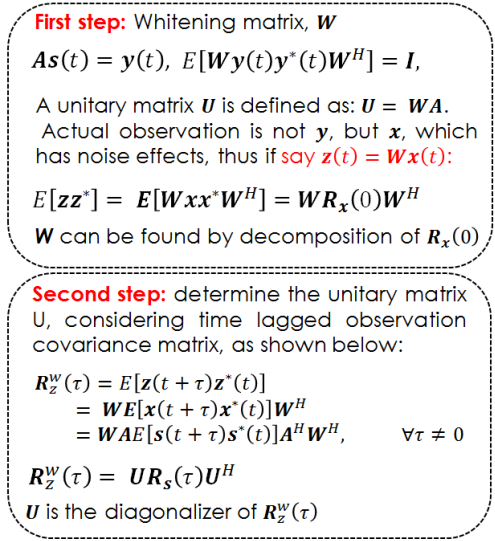}
\caption{SOBI process flowchart}
\label{SOBI_fc}
\end{figure}

Following the implementation of one of the methods above, the residual signal is a linear mixture of two sources, $y_k^{rgh}$ and $y_k^{brg}$, as shown in Equation \ref{eq:2}. At this point, the goal is to extract the bridge response from this signal to enable SID. BSS techniques for linear/static mixtures are thoroughly studied in different scenarios \cite{sadhu2017review,cichocki2002adaptive,choi2005blind,hyvarinen2000independent}. Second order blind identification (SOBI) \cite{tong1990amuse,belouchrani1997blind} is a successful BSS technique for separation of linear mixtures, especially for spectrally uncorrelated sources. The general mathematical description of a linear mixture is demonstrated in Equation \ref{eq:SOBI}.

\begin{equation}\label{eq:SOBI}
x(t) = As(t)+ \sigma(t) = y(t)+\sigma(t)
\end{equation}

where $A$ is a constant mixing matrix, $s(t)$ is a vector matrix of sources and $x(t)$ is a vector matrix of mixed signals. $\sigma(t)$ is also the additive noise in the observations. SOBI assume source uncorrelatedness, i.e., $R_s(\tau)$ is diagonal. In addition, for simplicity, the covariance matrix of the sources is presumed to be an identity matrix. The mathematical process is shown at the flowchart presented in Figure \ref{SOBI_fc}. In the first step, a whitening matrix $W$ should be calculated to diagonalize the observation covariance matrix. In the second phase, unitary matrix U shall be found to satisfy $R_z^w(\tau)=UR_x(\tau)R_z^w(\tau)$. Given $W$ and $U$, the mixing matrix can be found as $A = W^{-1}U$. 

In the SOBI formulation, $\tau$ is an arbitrary time-lag. For a more robust solution, a set of time-lags is recommended \cite{belouchrani1997blind} to be considered for diagonalization. In this manner, the unitary matrix U that maximally diagonalizes all time-lagged covariance matrices in the set, is selected. In this study, SOBI is applied to separate the bridge vibrations from the road surface effects so that STRIDEX can be implemented as illustrated in Figure \ref{methods_fc}. 


\section{Generation of Vehicle Scanning Data} \label{sec:vcl_sim}

\subsection{Bridge finite element analysis}
\label{S:1}

\begin{figure}[!ht]
\begin{subfigure}{1.0\textwidth}
\centering\includegraphics[width=0.7\linewidth]{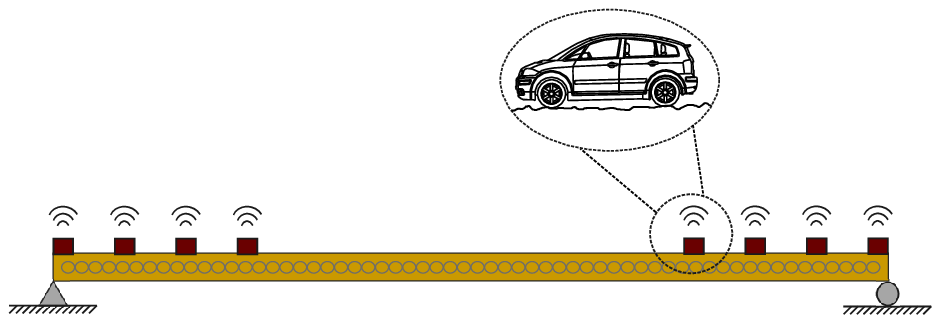}
\caption{side view of the sensing scenario}
\end{subfigure}\\
\begin{subfigure}{1.0\textwidth}
\centering\includegraphics[width=0.7\linewidth]{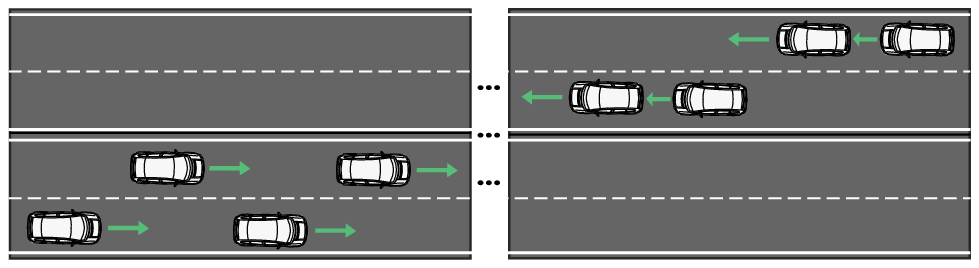}
\caption{Illustrative plan of the sensing scenario}
\end{subfigure}\\
\caption{Sensing vehicles layout}
\label{simul_layout}
\end{figure}

A $500 m$ long bridge with rigid constraints at both ends is studied to evaluate the performance of the proposed methods. The criteria for selecting the bridge model are as follows: (1) maximum consistency in modal characteristics with the real-world long-span bridges, (2) simplification of the analysis, and (3) generalization of the model. The goal is to identify modal properties of this bridge from signals collected by a mobile sensor network comprised of eight vehicles as shown in Figure \ref{simul_layout}. The bridge is numerically modeled with elastic beam elements with $1728$Kg nodal mass at each DOF, $17.28m^2$ cross-sectional area, and $85.81m^4$ moment of inertia. These characteristics are set in a way that the bridge shows realistic modal properties (i.e., operational natural frequencies, mode shapes, and damping ratios) compared to existing bridges \cite{abdel1991importance,weng2008output}. Therefore, the model yields four natural modes at $0.1357$Hz, $0.3714$Hz, $0.7213$Hz, and $1.1710$Hz for the bridge. Simultaneously, three different road surface cases are analyzed: sinusoidal, expansion joints, and random. The bridge is excited by random white noise acting on nine point equally spaced along the bridge to simulate collective effect of random ongoing traffic. Numerical analysis of the bridge is performed in $OpenSees$ finite element (FE) program \cite{mckenna2000open}. The damping for the first and sixth modes are set as $2\%$ using Rayleigh's method. Numerically, the model has been discretized into $10,000$ degrees of freedom, so will be called as $10K$ DOF model. As demonstrated in Figure \ref{simul_layout}, the sensing scenario consists of eight vehicles, each traveling $70\%$ of the bridge span. Half of the vehicles travel right-to-left and the other half go the opposite. The vehicles travel with $2.5 \frac{m}{s}$ velocity. This low speed has set in order to minimize excitations of the vehicle suspension system, as suggested in \cite{siringoringo2012estimating, yang2009extracting}. Note that the vehicles shown in Figure \ref{simul_layout} are located in different lanes to demonstrate the vehicle layout possibilities, however, since the 2D model of the bridge is considered, torsional modes are not estimated in this study. In these figures (and many other figures in this paper), the x-axis is labeled by the number of the DOF within the length of the bridge from the numerical model. In fact, DOFs associate with different locations over the modeled bridge. \par

\begin{figure}[!ht]
\centering\includegraphics[width=1.0\linewidth]{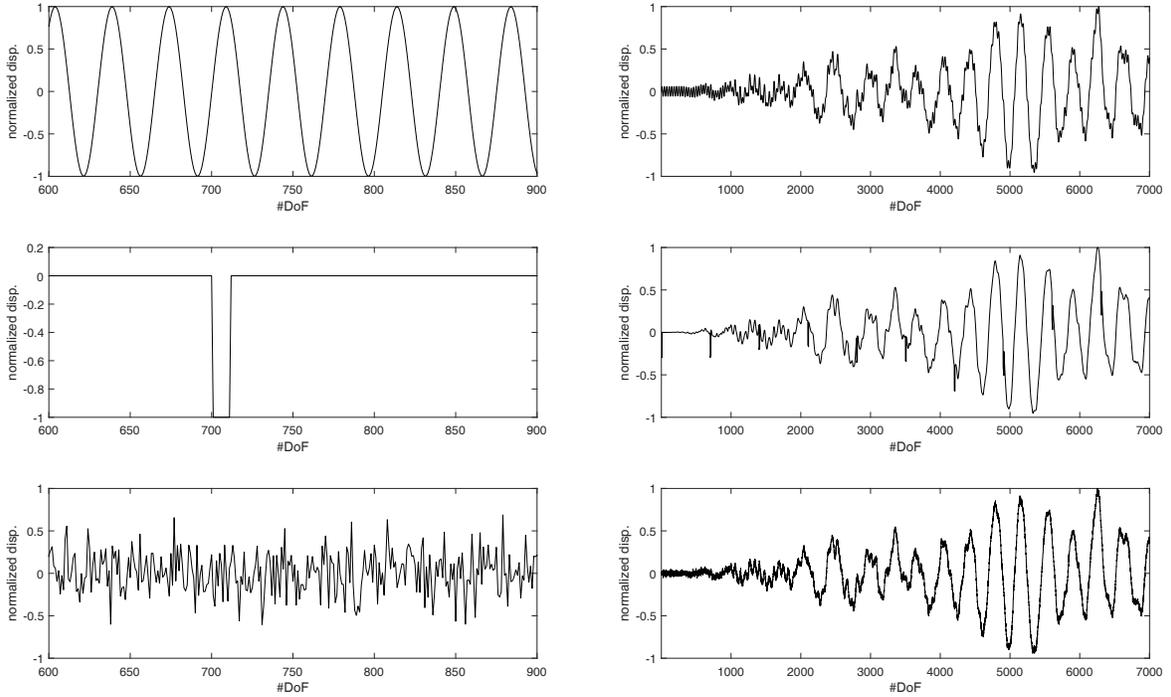}
\caption{Left) Road roughness profile, Right) Total displacement felt by tires. From top to bottom: sinusoidal, expansion joints, and random white noise roughness profiles.}
\label{vcl_input}
\end{figure}

In order to produce signals collected by vehicles from the vehicle-bridge interaction (VBI), the approach is to first, produce vehicle input as shown in Equation \ref{eq:2} and then, pass the input through a two DOF vehicle quarter-car model to generate its response. In Equation \ref{eq:2}, the vehicle input contains two components: bridge response, and roughness displacement at vehicle locations. For the bridge part, at the end of the FE analysis, a dense matrix of the bridge response at all 10K DOFs and time steps is computed. Next, for each vehicle, the values of bridge response at some DOFs and times are collected in accordance to the moving vehicle locations at any time. After finding a vector of the bridge response according to the location, corresponding roughness displacements are added location-wise to complete vehicle input generation phase. Once the compounded input signal is generated, the vehicle is subjected to it for vehicle response simulation. Figure \ref{vcl_input} shows a displacement time signal felt by one of the vehicles while traveling over different types of road profiles. 

\begin{figure}[!ht]
\centering\includegraphics[width=1.0\linewidth]{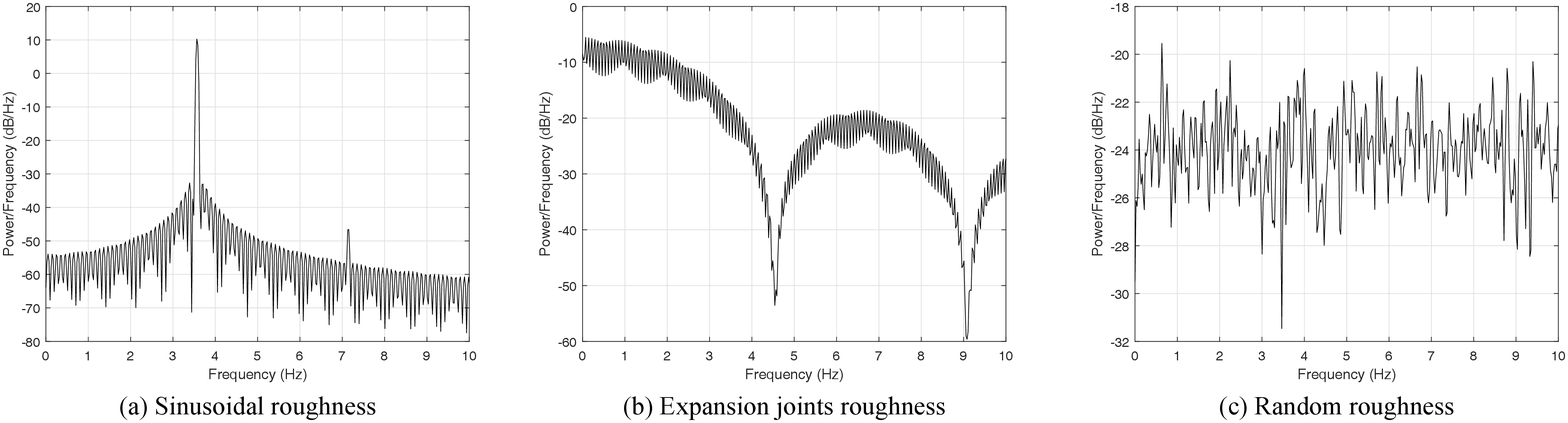}
\caption{DFT of roughness cases - (b) indicates high amplitude fluctuations, will be propagated later}
\label{rgh_dft}
\end{figure}

Three roughness patterns are investigated to examine robustness of the suggested methods. The first is a simple sinusoidal function; the second   is a random white noise sequence, and the last is an expansion joint (EJ) model consisting of a white noise signal with a series of impulses (0.5 meter wide drops). The sinusoidal roughness is selected due to its simplicity in frequency domain, to verify the methods plausibility. The second case is an estimation of a real pavement profile and the latest is designed to simulate bridge expansion joints. This case also can be a representation of any possible road obstacles, such as potholes or speed bumps. In the following sections, the third pattern will be noted as EJ for brevity. The Welch's power spectral density (PSD) functions of the roughness profiles are also shown in Figure \ref{rgh_dft}. The PSD of sinusoidal roughness shows a single spike, while two other cases have more complex frequency representations. Note that in the PSD of the expansion joints case, high energy fluctuations are superimposed over a relatively smooth hill series. This effect is intrinsic caused by the sudden drops in the time signal, and will be propagated and observable later. \par 

Note that in Figure \ref{simul_layout} the mobile sensors are synchronized and start collecting data simultaneously. In addition, roughness extraction using SOBI needs at least two channels of data from the same portion of the bridge scanned by each vehicle to produce two sources. These signals do not need to be synced. As a result, at least one addition mobile sensing data for each vehicle path is needed. These additional measurements can be collected before or after the main experiment (synchronized data collection of eight vehicle, Figure \ref{simul_layout}), resulting minimal operational difficulties. 

\subsection{Simulation of the vehicle system}
\label{S:2}

After generating the displacement input of the vehicles, these vectors are applied to a quarter-car model of the vehicle, which is a simplified and common model of the vehicular suspension system. Two factors are important for vehicle properties: frequency band separation between vehicle and the bridge, and relatively low damping for the vehicle. The first is a common preference for BSS methods, such as EEMD. In fact, as per \citet{rilling2008one}, EEMD is not able to separate harmonics with frequency ratio of greater than $0.6$ (small frequency over the large frequency). The second criterion also enables the SID of the vehicle in the preprocessing phase. Qualitatively, for long-span bridges, stiff cars and for short bridges, flexible cars are appropriate choices in terms of frequency bands separation, For example, for a bridge with frequencies in the range of 8 - 25 Hz inspected by a vehicle with a frequency of 4 Hz or below, EEMD would successfully separate the vehicle dynamics. To demonstrate this approach, a vehicle with properties shown in Table \ref{table:1} is considered. The dynamic response of the vehicle is assumed to remain linear throughout data collection. In the following section, another vehicle, with more common dynamical properties is analyzed. Note that these examples are generated using a generic vehicle model and are presented to validate the proposed methods; this is not a comprehensive report on how vehicle parameters influence identification results. The quarter-car suspension model is attached to the ground via a point, meaning that the tire touches every samples of the displacement input signal. 

\begin{table}[!htbp]
\centering
\caption{Vehicle properties}
\begin{tabular}{l l l}
\hline
\textbf{Property Name} & \textbf{Value} & \textbf{Units}\\
\hline
Unsprung Mass & 49.8 & Kg \\
Sprung Mass & 466.5 & Kg \\
Tire Damping & 0.0 & Ns/m \\
Suspension Damping & 1400 & Ns/m \\
Tire Stiffness & 720 & kN/m \\
Suspension Stiffness & 1,800 & kN/m \\
Fundamental Frequency & 5.14 & Hz \\
\hline
\end{tabular}
\label{table:1}
\end{table}

\begin{figure}[!ht]
\centering\includegraphics[width=1.0\linewidth]{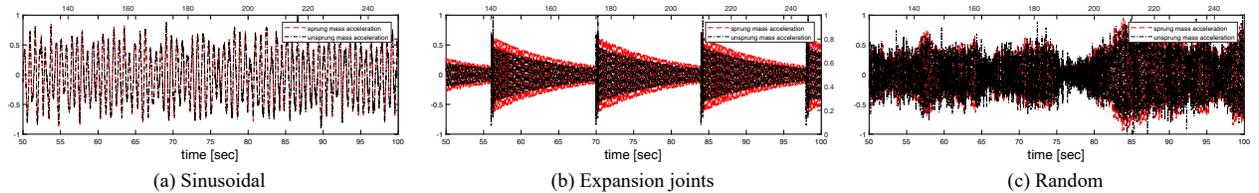}
\caption{Vehicle normalized outputs resulted from simulation}
\label{simul_res}
\end{figure}

The quarter-car model for the vehicle is simulated using a state-space model with inputs linked to the road profiles as shown in Figure \ref{vcl_input}. The simulation outputs for three roughness profiles are shown in Figure \ref{simul_res} (plots show vehicle responses at both DOF's, tire and cabin levels). 

The figure shows that the vehicle response of the vehicle under identical bridge load cases, is highly sensitive to the road roughness profile. In the next step, these vehicle responses are used as the input for the separation phases and will be processed further to extract bridge modal properties.

\section{Extraction of Bridge Vibrations from Vehicle Scanning Data}\label{sec:brg_ext}
\subsection{Method 1 - Signal deconvolution with FRF}

In general, the problem can be revisited as two cascaded blocks; first a bridge which is responding linearly to ambient random loads; the second, a vehicle which is excited by both the instantaneous vertical vibrations of the bridge as well as the road surface roughness, as illustrated in Figures \ref{prb_def} and \ref{simul_layout}. For vehicle simulation, the properties shown in Table \ref{table:1} has been introduced. The objective in this section is to remove the effect of the vehicle suspension system on the signal and retrieve the vehicle input, which is a linear mixture of other sources. As the most direct approach, deconvolution of vehicle output using vehicle frequency response function is examined. FRF is a function of system dynamic properties and has to be identified in advance, knowing that the existing conditions and properties of vehicles are different. With this in mind, a preprocessing phase is adopted to identify the vehicle using output-only SID methods. 

\subsubsection{Vehicle system identification and frequency response function}

\begin{figure}[!htbp]
\begin{subfigure}{.6\textwidth}
\centering\includegraphics[width=0.8\linewidth]{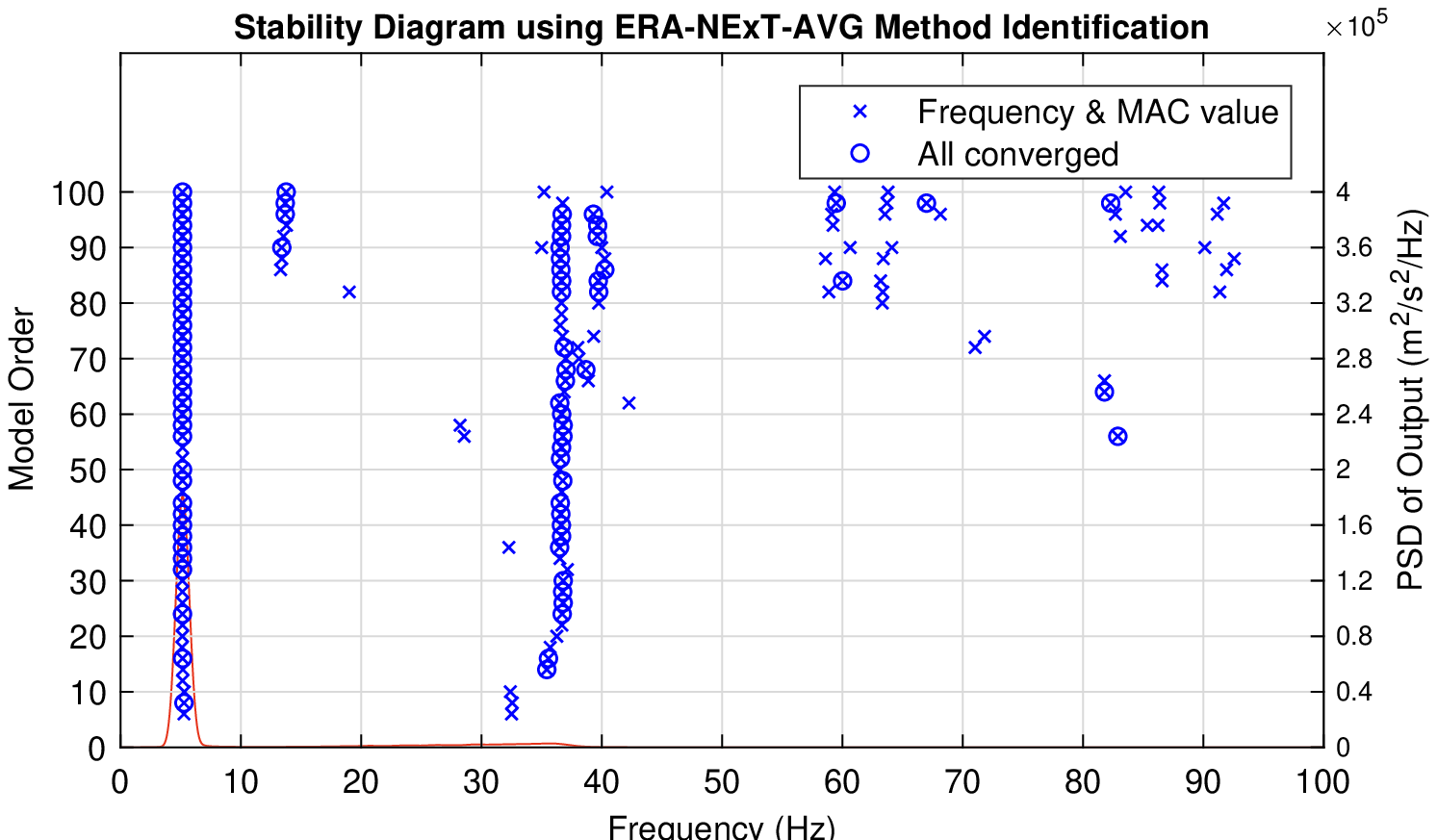}
\end{subfigure}%
\begin{subfigure}{.4\textwidth}
\centering\includegraphics[width=1.0\linewidth]{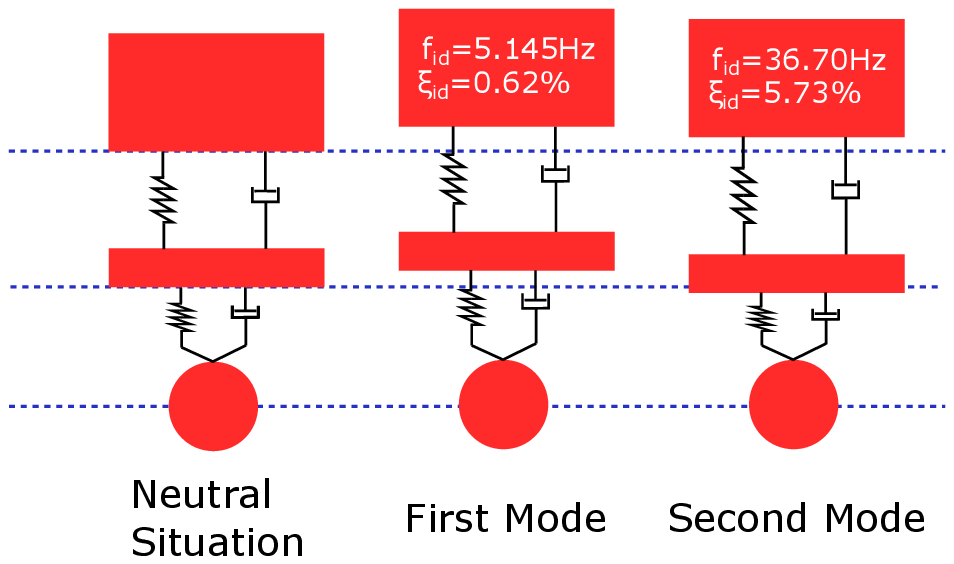}
\end{subfigure}
\caption{vehicle identification results - SMIT \citep{SMIT} package outputs}
\label{vcl_SID}
\end{figure}

Output-only system identification is a well-studied field and there are many successful methods that can identify linear systems under ambient random loads. A toolsuite is developed by \citet{SMIT} that has integrated some of these algorithms for SID. In the application of vehicle SID, it has been assumed that the vehicle response is recorded at both degrees of freedom (tire level and the cabin) while it is passing over a rough pavement with a random Gaussian pattern. Figure \ref{vcl_SID} shows the results of vehicle identification using its ambient response under random roughness. Exact natural frequencies of the vehicle are 5.14 Hz and 36.78 Hz. The identified frequencies shown in the table match well with these natural frequencies.

\begin{figure}[!htbp]
\centering\includegraphics[width=0.5\linewidth]{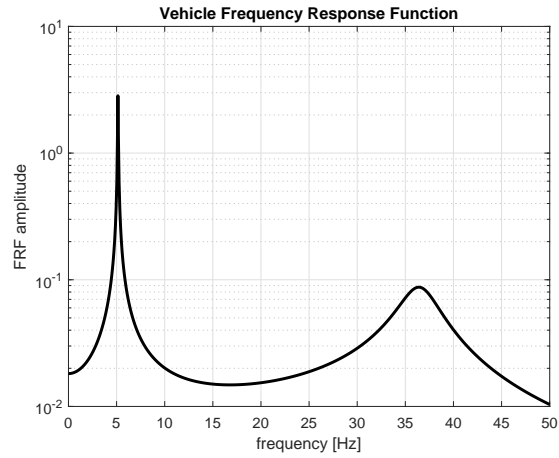}
\caption{vehicle FRF from inputs at tire level to response inside the room}
\label{vcl_FRF}
\end{figure}

The next step is to construct the vehicle FRF from identified characteristics found in the previous part. For this purpose, Equations \ref{eq:FRF2} and \ref{eq:FRF3} below are used \cite{bilovsova2011FRF}:

\begin{equation}\label{eq:FRF2}
\left[\boldmath{\alpha}(\omega)\right]=\left[K+i\omega C-\omega^2M\right]
\end{equation}
\begin{equation}\label{eq:FRF3}
\alpha_{jk}(\omega)=\sum_{r=1}^{N} \frac{\Phi_{rj}.\Phi_{rk}}{\Omega_r^2-\omega^2+2i\omega\Omega_r\xi_r}
\end{equation}

where $\alpha_{jk}$ is the FRF that maps an input load at DOF $j$ to the response at DOF $k$. $\Phi$'s are mode shapes and $\Omega_r$ is the undamped frequency of mode $r$. Equation \ref{eq:FRF3} produces the vehicle FRF which is useful to find the vehicle input from its response collected in the cabin (can reproduce the vehicle input). Using this equation, the corresponding FRF is generated to be used for deconvolution, shown in Figure \ref{vcl_FRF}. As pointed out, the first frequency spike happens at $5.142 Hz$ which is very close to the actual frequency of the car.

\begin{figure}[!ht]
\centering
\centering\includegraphics[width=1.0\linewidth]{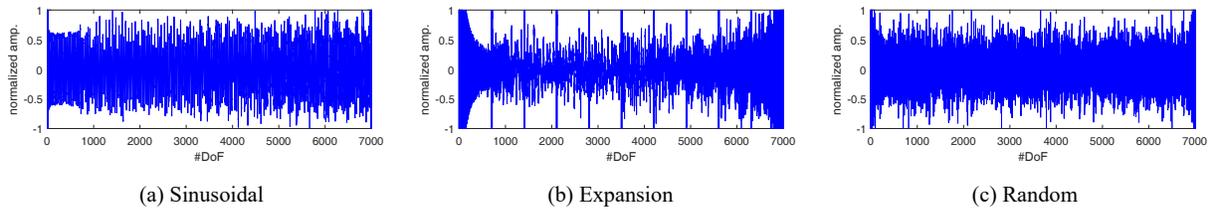}
\caption{Time signals after deconvolution}
\label{vcl_time_deconv}
\end{figure}

\begin{figure}[!ht]
    \centering\includegraphics[width=0.5\linewidth]
    {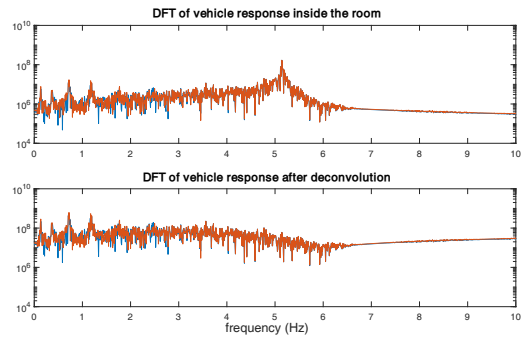}
	\caption{DFT of signals before and after deconvolution using FRF for random roughness}
	\label{vcl_deconv}
\end{figure}

Note that the transfer function can be derived in closed-form as well if the manufactured properties of the vehicle is available. For instance, \cite{sun2001modeling} and \cite{bogsjo2012models} have provided two sets of equations for the transfer function of the sprung and unsprung DOF's of a vehicle. Using the transfer function $H(\omega)$ (FRF), simulation outputs can be deconvolved by Equation \ref{eq:3} via element-wise division of the discrete Fourier transform (DFT). Figures \ref{vcl_time_deconv} and \ref{vcl_deconv} show the results of deconvolution using the identified FRF. 

Figure \ref{vcl_deconv} shows that after deconvolution, the vehicle frequency content has been perfectly removed (the sharp spike around $5Hz$). In addition, a comparison of time signals in Figures \ref{simul_res}b and \ref{vcl_time_deconv}b (and other cases with less clarity) shows that the impulse effects caused by roughness shocks in the vehicle response have also been discarded. The remaining signal is a linear mixture of two sources; bridge vibrations and roughness profile displacements. The next step is to apply second order blind identification (SOBI) to separate these sources and extract the bridge vibrations.

\subsection{Method 2 - Signal deconvolution with EEMD}

\begin{figure}[!htbp]
\centering\includegraphics[width=1.0\linewidth]{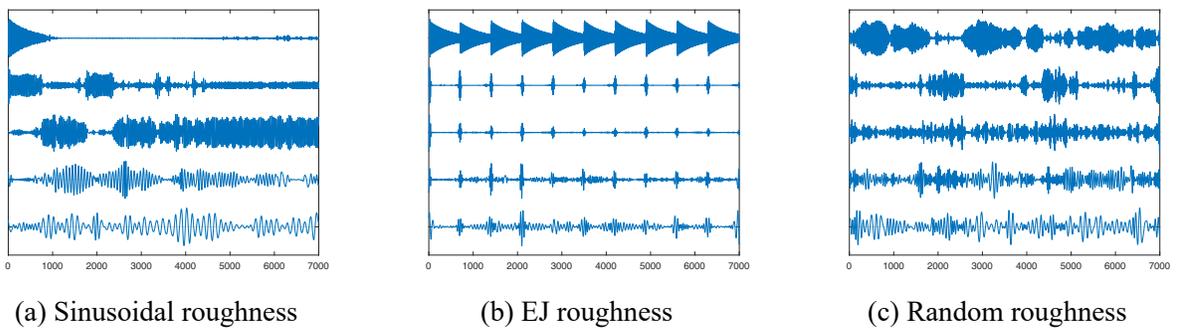}
\caption{IMFs in time domain}
\label{eemd_strips}
\end{figure}

\begin{figure}[!ht]
\centering\includegraphics[width=1.0\linewidth]{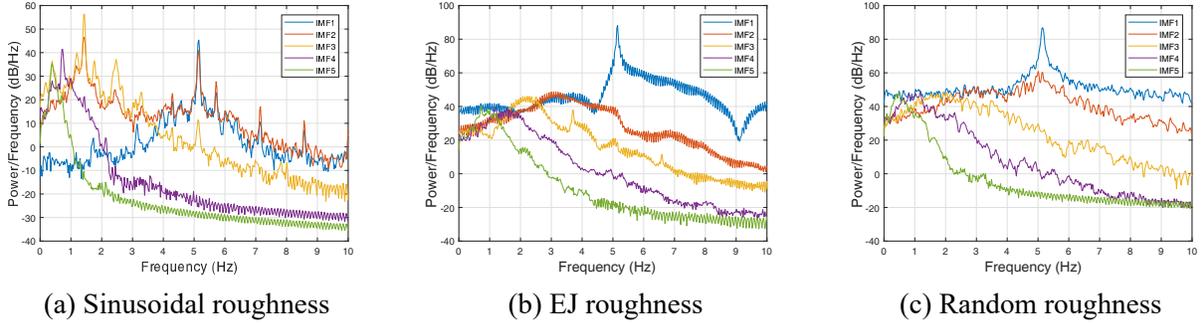}
\caption{IMFs in frequency domain}
\label{eemd_psd}
\end{figure}

\begin{figure}[!htbp]
\centering\includegraphics[width=1.0\linewidth]{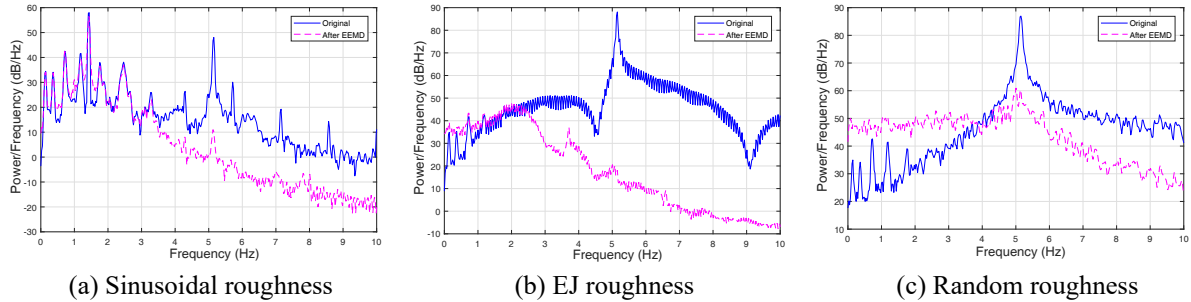}
\caption{Original signal before and after EEMD}
\label{psd_eemd_before_after}
\end{figure}

In this section, EEMD is applied to eliminate vehicle effects. Figure \ref{eemd_strips} shows the oscillatory components of the signal called intrinsic mode functions (IMF). IMFs are the building components of a mixed signal that are extracted by the EEMD algorithm \cite{huang1998empirical}. In summary, EEMD algorithm detects the most major building oscillation in a mixed signal by finding the envelope of the time signal. Then the major oscillation which is the first IMF is removed from the original signal and the process is repeated on the remaining signal to extract all IMFs. The frequency representation of IMFs are presented in Figure \ref{eemd_psd}. The first IMF is able to capture vehicle effect and is presented in both time and frequency domains. This vehicle extraction is ideal for the case with EJ roughness, shown in Figure \ref{eemd_psd}b, yet is not as perfect for two other cases. Figure \ref{eemd_psd}a and \ref{eemd_psd}c show traces of the vehicle response in their second IMF as well (frequency content around $5Hz$ is high). Therefore, in order to remove the vehicle effect from the signal, two first IMFs are subtracted from the original signal; the resultant signals are represented in Figure \ref{psd_eemd_before_after}. The deconvolution using EEMD could not perfectly remove the vehicle frequency content, as shown in Figure \ref{psd_eemd_before_after}, however, the remainder contains considerably less energy now. It will be seen in the following sections that despite this imperfect deconvolution, it is adequate for bridge modal identification purposes. 

Note that the method is easier to perform compared to FRF method, since in this approach, there is no need for a preprocessing step for identifying the vehicle in advance. In fact, this method gives an estimate of the pure vehicle response (which is equivalent to the vehicle identification) as one of the extracted IMFs. A drawback of this method is its inability to extract closely-spaced frequencies \cite{flandrin2004empirical}. However, in a general setting, the frequency contents of the vehicle and the bridge may overlap. Thus, while EEMD is advantageous in its ease of use, it is not a universal solution.

In this case, the spatial frequency contents of the roughness cases are located mostly within the same region as the bridge frequency content and EEMD was not able to extract them as a separate source because of its disability for extracting contents with closely-spaced frequencies. Thus, for roughness separation (regardless of vehicle separation method), an extra step is necessary.

\subsection{SOBI for linear un-mixing}

\begin{figure}[!htbp]
\centering\includegraphics[width=1.0\linewidth]{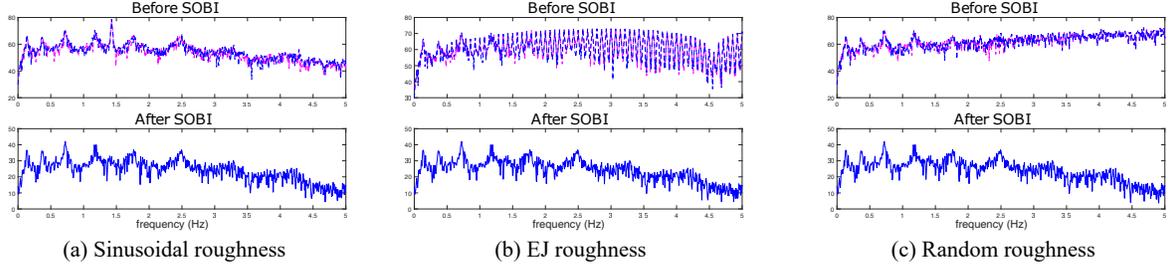}
\caption{SOBI on FRF output}
\label{eemd_frf}
\end{figure}

\begin{figure}[!ht]
\centering\includegraphics[width=1.0\linewidth]{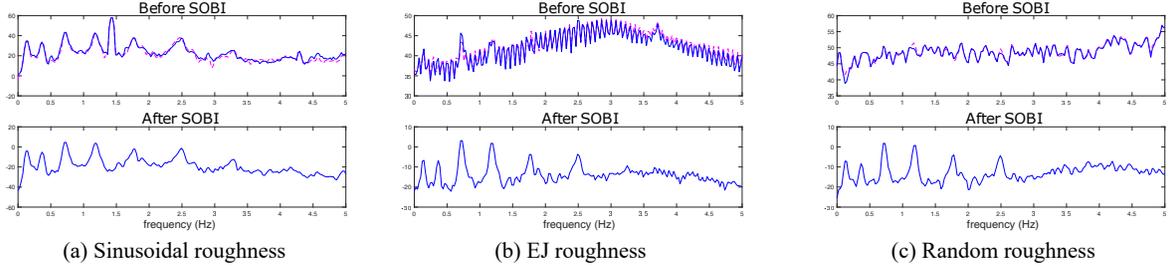}
\caption{SOBI on EEMD output}
\label{eemd_sobi}
\end{figure}

Second order blind identification (SOBI) \cite{poncelet2007output} is a method for unmixing linearly mixed signals. In this problem, the remaining sources, bridge vibrations and road profile roughness displacements are assumed to be unknown, however, some assumptions regarding them hold, such as being uncorrelated. The SOBI algorithm is implemented in \textregistered$MATLAB$ and used for separating roughness and bridge contents of the signals derived from deconvolution. Results of SOBI applied on outputs of both methods are shown in Figures \ref{eemd_frf} and \ref{eemd_sobi} (bottom plots show the extracted source corresponding to the bridge only). SOBI is successful in diminishing roughness-induced peaks from the signal in both methods. The remaining signal is bridge vibrations and can further be utilized for system identification using STRIDEX. Note that the PSD of the signal resulted from FRF (Figure \ref{eemd_frf}) has higher spectral resolution comparing to the other approach. This is because that in the FRF method, there is a possibility to enhance the deconvolution quality by refining the frequency range. However, it has been realized by authors that the technique is insensitive in the EEMD method.

To summarize, throughout these steps, the signal collected by a moving vehicle over a bridge has been processed to extract the bridge vibration signal. In the next step, these signals are analyzed by the system identification algorithm, STRIDEX, for bridge modal identification. Note that while the extraction approaches discussed were applied on the channels individually, STRIDEX can operate on multiple sensor channels simultaneously.

\section{Bridge Modal Identification}

In this simulated case, eight vehicles travel over a bridge modeled by 10K degrees of freedom. Each vehicle trip (one direction) scans 7K of DOFs in one direction as presented in Figure \ref{simul_layout}a. In this section, given extracted bridge dynamic vibration signals from mobile sensors, the procedure of STRIDEX \cite{matarazzo2018scalable} for bridge system identification is performed.

\begin{figure}[!ht]
\centering\includegraphics[width=0.9\linewidth]{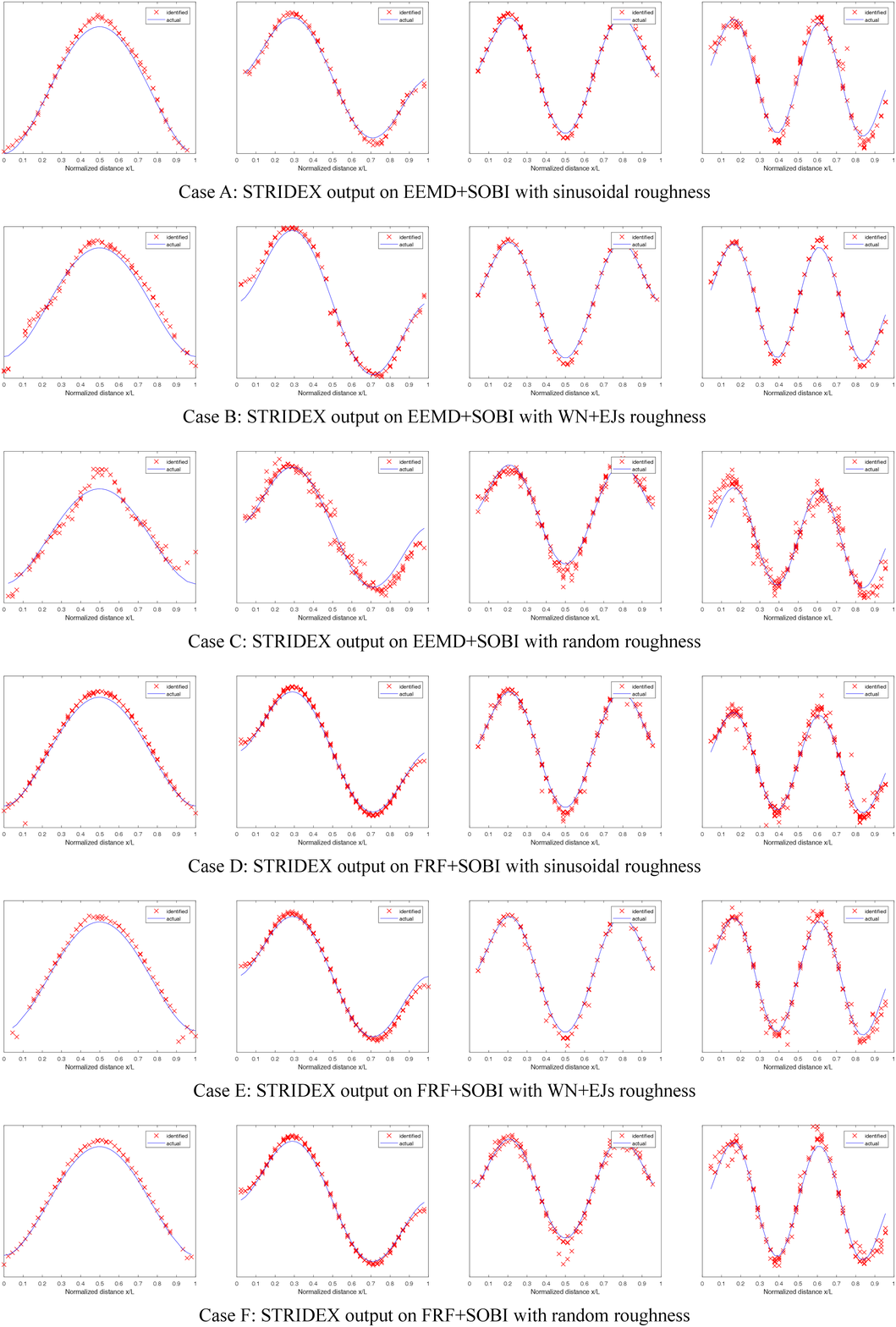}
\caption{Model identification results}
\label{stridex_1}
\end{figure}

The methods create many outputs among which, mode shapes were selected manually, which was a tedious task \cite{matarazzo2018scalable}. As a contribution of this study, an algorithm is proposed in Appendix \ref{append1:mode_aggregate} to do the process of mode selection and superposition in an automated way and is implemented here. Identified modal parameters from this algorithm are shown in Figure \ref{stridex_1}. 

\begin{table}[!ht]
\centering
\caption{Identified frequencies}
\label{iden_frq}
\begin{tabular}{cc|c|c|c|c|c|c|}
\cline{3-8}
\multicolumn{2}{c|}{\textit{values in Hz}}          & \multicolumn{3}{c|}{\textit{EEMD+SOBI}} & \multicolumn{3}{c|}{\textit{FRF+SOBI}} \\ \hline
\multicolumn{1}{|c|}{Mode ID}    & \textbf{Actual} & Sinusoidal     & EJ         & Random    & Sinusoidal     & EJ        & Random    \\ \hline
\multicolumn{1}{|c|}{\textbf{1}} & \textbf{0.1357} & 0.1345         & 0.1343     & 0.1366    & 0.1360         & 0.1354    & 0.1361    \\ \hline
\multicolumn{1}{|c|}{\textbf{2}} & \textbf{0.3714} & 0.3689         & 0.3689     & 0.3677    & 0.3691         & 0.3687    & 0.3689 \\ \hline
\multicolumn{1}{|c|}{\textbf{3}} & \textbf{0.7213} & 0.7218         & 0.7195     & 0.7204    & 0.7214         & 0.7227    & 0.7204 \\ \hline
\multicolumn{1}{|c|}{\textbf{4}} & \textbf{1.1710} & 1.1832         & 1.1804     & 1.1861    & 1.1836         & 1.1850    & 1.1782 \\ \hline
\end{tabular}
\end{table}

\begin{table}[!ht]
\centering
\caption{Identified damping ratios}
\label{iden_dmp}
\begin{tabular}{cc|c|c|c|c|c|c|}
\cline{3-8}
\multicolumn{2}{c|}{\textit{values in \%}}          & \multicolumn{3}{c|}{\textit{EEMD+SOBI}} & \multicolumn{3}{c|}{\textit{FRF+SOBI}} \\ \hline
\multicolumn{1}{|c|}{Mode ID}    & \textbf{Actual} & Sinusoidal     & EJ        & Random     & Sinusoidal     & EJ       & Random     \\ \hline
\multicolumn{1}{|c|}{\textbf{1}} & \textbf{2.00}   & 3.26           & 1.49      & 2.04     	 & 4.30           & 3.91     & 4.18 \\ \hline
\multicolumn{1}{|c|}{\textbf{2}} & \textbf{0.80}   & 0.92           & 0.76      & 0.94       & 0.82           & 0.78     & 0.84 \\ \hline
\multicolumn{1}{|c|}{\textbf{3}} & \textbf{0.60}   & 0.69 			& 0.55      & 0.14       & 1.59           & 2.00     & 0.70 \\ \hline
\multicolumn{1}{|c|}{\textbf{4}} & \textbf{0.70}   & 0.95           & 0.32      & 0.35       & 0.90           & 1.46     & 0.31 \\ \hline
\end{tabular}
\end{table}

\begin{table}[!ht]
\centering
\caption{Identified modal assurance criteria (MAC)}
\label{iden_MAC}
\begin{tabular}{c|c|c|c|c|c|c|}
\cline{2-7}
\textit{values in \%}            & \multicolumn{3}{c|}{EEMD+SOBI} & \multicolumn{3}{c|}{FRF+SOBI} \\ \hline
\multicolumn{1}{|c|}{Mode ID}    & Sinusoidal  & EJ     & Random  & Sinusoidal  & EJ     & Random \\ \hline
\multicolumn{1}{|c|}{\textbf{1}} & 0.9991       & 0.9944  & 0.9829   & 0.9997       & 0.9996  & 0.9997  \\ \hline
\multicolumn{1}{|c|}{\textbf{2}} & 0.9914       & 0.9856  & 0.9640   & 0.9942       & 0.9922  & 0.9934  \\ \hline
\multicolumn{1}{|c|}{\textbf{3}} & 0.9825       & 0.9806  & 0.9612   & 0.9806       & 0.9816  & 0.9689  \\ \hline
\multicolumn{1}{|c|}{\textbf{4}} & 0.9605       & 0.9802  & 0.9443   & 0.9628       & 0.9612  & 0.9536  \\ \hline
\end{tabular}
\end{table}

\subsection{Modal property results}
Tables \ref{iden_frq}, \ref{iden_dmp} and \ref{iden_MAC} show natural frequencies and damping ratios comparing to the actual values, and modal assurance criteria (MAC) values respectively. MAC is an indicator of fitting accuracy between estimated mode shapes and actual ones \cite{pastor2012modal}. It is demonstrated in Table \ref{iden_frq} and Figure \ref{stridex_1} that both proposed methods are successful in estimating natural frequencies and mode shapes of the bridge. As Figure \ref{stridex_1} case C demonstrates, in the case of random roughness, the first mode is not ideally obtained, however, FRF method has extracted four modes from data collected over this road condition. Table \ref{iden_dmp} shows that the damping values are estimated precisely in most cases. Two methods (EEMD and FRF) are relatively as successful in damping estimation. In the case of the first mode identified with FRF method, damping values are not as close as others with respect to the exact values. All estimated modes have MAC value within $0.9443$ to $0.9997$ range (Table \ref{iden_MAC}), which indicate the methods strength. In terms of mode shape accuracy, a comparison between methods shows that the FRF method outperformed slightly. Note that the identified MAC values are dependent to the threshold set for the automated mode aggregation algorithm (Appendix \ref{append1:mode_aggregate}); if lower threshold is selected, higher MAC can be found, however, the identified modes would have less resolution. \par

While the primary goal of this paper is bridge modal identification using mobile sensor measurements, the proposed methods are able to estimate other important characteristics of the problem, e.g., the road surface roughness or the vehicle dynamical properties. These capabilities are discussed and presented in Appendix \ref{append2:auxiliary}. 

\section{Application using Mechanical Properties of Commercial Vehicles}\label{comm_car}

Regular vehicles commonly have a frequency range between 0.5 Hz to 1.5 Hz, as recommended by Olley Criteria \cite{milliken2002chassis} for a comfortable ride. This frequency range is highly possible to overlap with bridge frequency band. In fact, some special types of automobiles, e.g., heavy trucks, bicycles and sports cars \cite{giaraffa2017tech,esmailzadeh1995ride,champoux2007bicycle} have high fundamental frequency close to the one derived from Table \ref{table:1}. The proposed methods work best when the vehicle's frequency do not overlap with the bridge frequency band of interest. The vehicle with mechanical properties shown in Table \ref{table:1} has a fundamental mode with about $5$ Hz frequency. Therefore, such a stiff vehicle is perfectly suitable for sensing long and flexible bridges with first few natural frequencies below $3$ Hz. In contrast, the fundamental frequency of a commercial car is about $1.0$ Hz which is a good match for shorter and stiffer bridges with fundamental frequency above $2.0$ Hz. In this section, a more common vehicle property set as shown in Table \ref{table:2} is considered to investigate the case of an overlapping frequency (The properties are scaled for a unit sprung mass). Note that in this study, the scope considers flexible bridges with stiff and commercial vehicles to emphasize the impact of various road profiles as well as vehicle types on bridge modal identification. However, the generality of the methods with respect to the bridge length has also been verified by considering shorter bridges (e.g. Appendix \ref{append0:short_bridge}). By eigenvalue analysis of the vehicle properties shown in Table \ref{table:2}, the natural frequencies are calculated as $1.64Hz$ and $11.01Hz$. For the brevity, the analyses plots are only shown for the case of random roughness pattern. However, identification results for all three cases are illustrated.  

\begin{table}[!ht]
\centering
\caption{Common vehicle properties}
\begin{tabular}{l l l}
\hline
\textbf{Property Name} & \textbf{Value} & \textbf{Units}\\
\hline
Unsprung Mass & 0.162 & Kg \\
Sprung Mass & 1.0 & Kg \\
Tire Damping & 0.0 & Ns/m \\
Suspension Damping & 1.86 & Ns/m \\
Tire Stiffness & 643 & N/m \\
Suspension Stiffness & 128.7 & N/m \\
Fundamental Frequency & 1.6 & Hz \\
\hline
\end{tabular}
\label{table:2}
\end{table}

\begin{figure}[!ht]
\centering\includegraphics[width=0.5\linewidth]{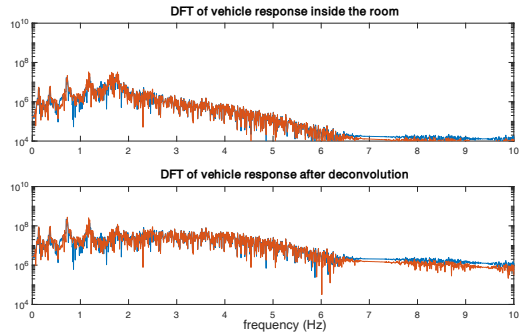}
\caption{FRF deconvolution on common vehicle}
\label{realvcl_deconv}
\end{figure}

\begin{figure}[!ht]
\centering\includegraphics[width=1.0\linewidth]{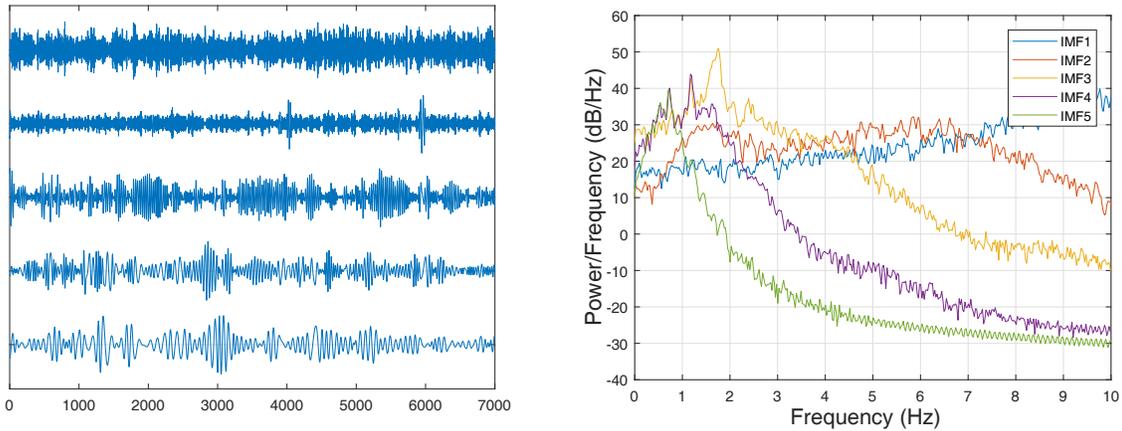}
\caption{EEMD results on common vehicle}
\label{realvcl_IMF}
\end{figure}

\begin{figure}[!ht]
\centering\includegraphics[width=1.0\linewidth]{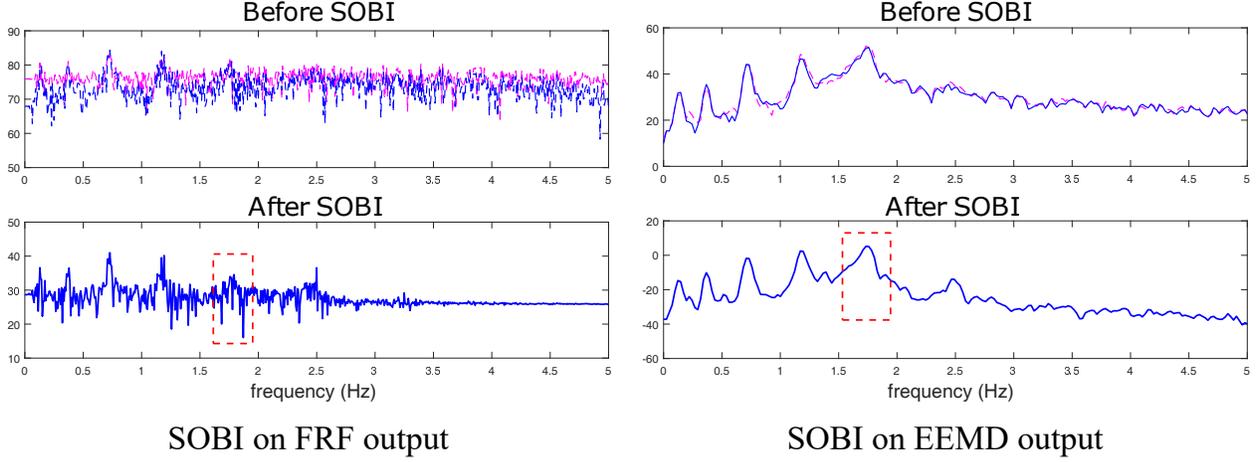}
\caption{SOBI on common vehicle}
\label{realvcl_SOBI}
\end{figure}

\begin{figure}[!ht]
\centering\includegraphics[width=0.9\linewidth]{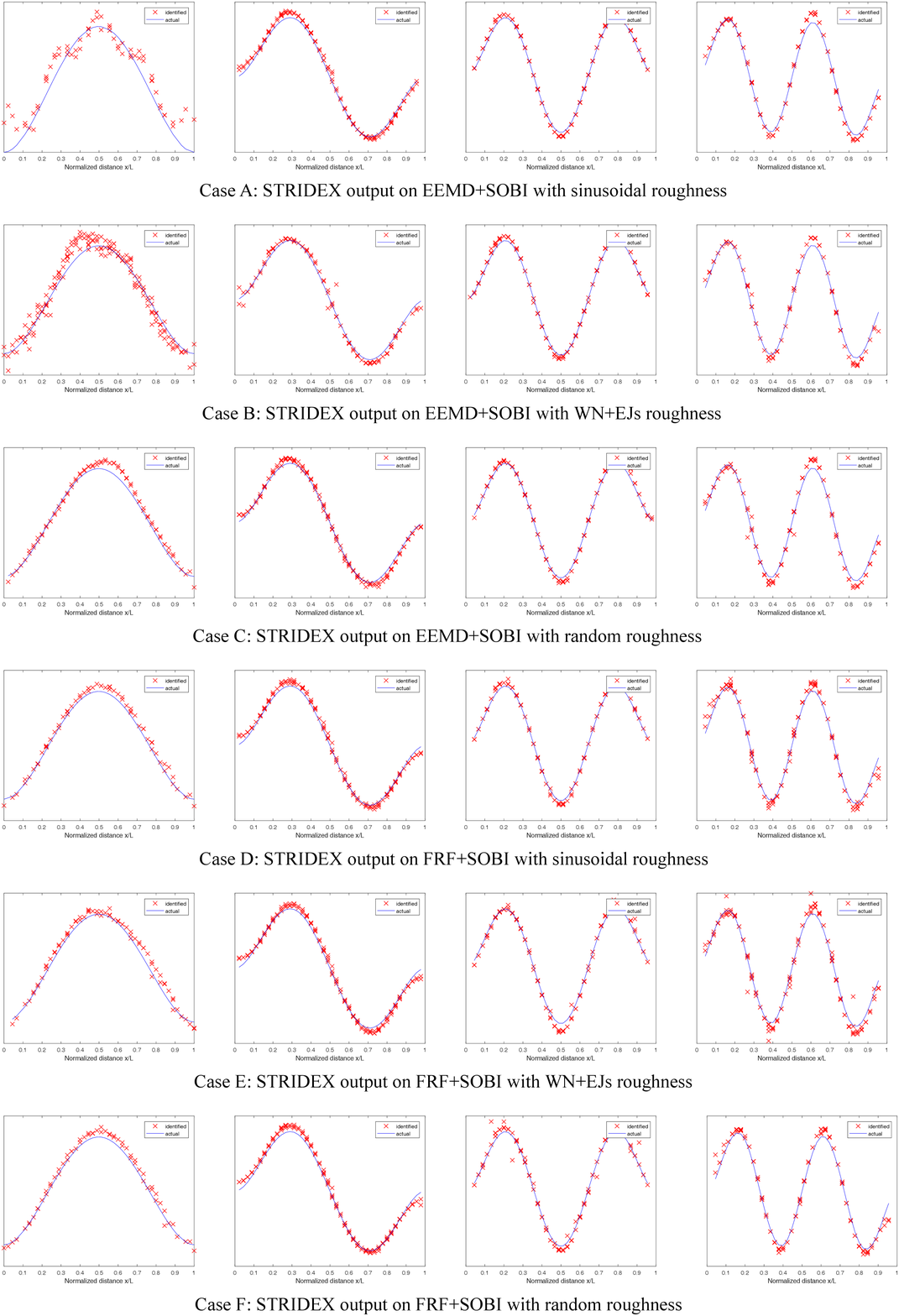}
\caption{Modal identification results - common vehicle}
\label{stridex_2}
\end{figure}

First, the vehicle deconvolution is performed. In the FRF method (Method 1), the frequency representation of the signal before and after deconvolution is shown in Figure \ref{realvcl_deconv}. Note that the vehicle transfer function was identified using output-only methods, as explained before. A similar task is done in EEMD (Method 2) by subtracting corresponding IMFs from the original signal to remove vehicle effects. EEMD IMFs are presented in both time and frequency in Figure \ref{realvcl_IMF}. The second IMF shows the vehicle content at $1.64Hz$, and is removed from the original signal. The next step is to extract roughness-induced vibrations using SOBI. Results for both methods are shown in Figure \ref{realvcl_SOBI} (Top plots show SOBI inputs, while bottom ones show the extracted bridge vibration source). Boxes in the plots are pointing to the fundamental frequency of the vehicle. Since the bands are closely spaced, both methods were not completely able to remove the vehicle content. However, the content is significantly weakened and is suitable for the bridge identification purposes, as shown in Figure \ref{stridex_2} and Tables \ref{iden_frq_1Hz} to \ref{iden_MAC_1Hz}. 

The identification results show accurate estimations of the mode shapes and natural frequencies, promising that the suggested methods are suitable for more general types of vehicles. A comparison between the results of this vehicle and the previous one indicate that as expected, the frequencies were estimated more accurately in the previous case. In particular, the identified properties of the first mode (frequency, damping ratio, and MAC value) are less desirable. An explanation for this observation is that since the energy carried by the first mode is less than others (lower power compared to other peaks in PSD plots (Figure \ref{realvcl_SOBI})), it is affected more by the vehicle contamination. The range of MAC values in Table \ref{iden_MAC_1Hz} is $0.9656$ to $0.9987$ which is desirable. By comparing Figure \ref{eemd_sobi} and Figure \ref{realvcl_SOBI}, it is noticeable that the first peak is wider in the second case. This wideness results in very high estimated damping ratios for the first mode, as shown in Table \ref{iden_dmp_1Hz}. The relatively lower accuracy of this case compared to the previous case is originated as a result of the frequency band proximity issue and was expected. \par

\begin{table}[!htbp]
\centering
\caption{Identified frequencies using common vehicle}
\label{iden_frq_1Hz}
\begin{tabular}{cc|c|c|c|c|c|c|}
\cline{3-8}
\multicolumn{2}{c|}{\textit{values in Hz}}          & \multicolumn{3}{c|}{\textit{EEMD+SOBI}} & \multicolumn{3}{c|}{\textit{FRF+SOBI}} \\ \hline
\multicolumn{1}{|c|}{Mode ID}    & \textbf{Actual} & Sinusoidal     & EJ         & Random    & Sinusoidal     & EJ        & Random    \\ \hline
\multicolumn{1}{|c|}{\textbf{1}} & \textbf{0.1357} & 0.1286         & 0.1250     & 0.1381    & 0.1423         & 0.1372    & 0.1404	 \\ \hline
\multicolumn{1}{|c|}{\textbf{2}} & \textbf{0.3714} & 0.3686 		& 0.3684     & 0.3688    & 0.3696         & 0.3696    & 0.3695    \\ \hline
\multicolumn{1}{|c|}{\textbf{3}} & \textbf{0.7213} & 0.7213 		& 0.7213     & 0.7207    & 0.7196         & 0.7209    & 0.7211    \\ \hline
\multicolumn{1}{|c|}{\textbf{4}} & \textbf{1.1710} & 1.1800 		& 1.1809     & 1.1801    & 1.1791         & 1.1923    & 1.1800	 \\ \hline
\end{tabular}
\end{table}

\begin{table}[!htbp]
\centering
\caption{Identified damping ratios using common vehicle}
\label{iden_dmp_1Hz}
\begin{tabular}{cc|c|c|c|c|c|c|}
\cline{3-8}
\multicolumn{2}{c|}{\textit{values in \%}}          & \multicolumn{3}{c|}{\textit{EEMD+SOBI}} & \multicolumn{3}{c|}{\textit{FRF+SOBI}} \\ \hline
\multicolumn{1}{|c|}{Mode ID}    & \textbf{Actual} & Sinusoidal     & EJ        & Random     & Sinusoidal     & EJ       & Random     \\ \hline
\multicolumn{1}{|c|}{\textbf{1}} & \textbf{2.00}   & 14.44 			& N.A.     	& 7.14       & 15.85     	  & 14.36    & 14.95       \\ \hline
\multicolumn{1}{|c|}{\textbf{2}} & \textbf{0.80}   & 0.70           & 0.82      & 0.40       & 1.01           & 1.43     & 1.02 		\\ \hline
\multicolumn{1}{|c|}{\textbf{3}} & \textbf{0.60}   & 0.47 			& 0.72      & 0.29       & 0.45           & 0.77     & 0.54       \\ \hline
\multicolumn{1}{|c|}{\textbf{4}} & \textbf{0.70}   & 0.37 			& 0.46      & 0.51       & 0.33           & 1.11     & 0.38       \\ \hline
\end{tabular}
\end{table}

\begin{table}[!htbp]
\centering
\caption{Identified modal assurance criteria (MAC) using common vehicle}
\label{iden_MAC_1Hz}
\begin{tabular}{c|c|c|c|c|c|c|}
\cline{2-7}
\textit{values in \%}            & \multicolumn{3}{c|}{EEMD+SOBI} & \multicolumn{3}{c|}{FRF+SOBI} \\ \hline
\multicolumn{1}{|c|}{Mode ID}    & Sinusoidal  & EJ     & Random  & Sinusoidal  & EJ     & Random \\ \hline
\multicolumn{1}{|c|}{\textbf{1}} & 0.9724       & 0.9883  & 0.9977   & 0.9987       & 0.9969  & 0.9987  \\ \hline
\multicolumn{1}{|c|}{\textbf{2}} & 0.9924       & 0.9921  & 0.9933   & 0.9920       & 0.9918  & 0.9909  \\ \hline
\multicolumn{1}{|c|}{\textbf{3}} & 0.9836       & 0.9822  & 0.9812   & 0.9836       & 0.9783  & 0.9726  \\ \hline
\multicolumn{1}{|c|}{\textbf{4}} & 0.9769       & 0.9779  & 0.9760   & 0.9743       & 0.9709  & 0.9656  \\ \hline
\end{tabular}
\end{table}

\section{System Identification using Approximated Vehicle Transfer Function}\label{sec:TF_est}

As a preprocessing part of the proposed method using FRF, sensing vehicles have to be identified in advance, as shown in Figure \ref{methods_fc}. For this identification, a complete description of the vehicle is needed for producing vehicle transfer function, hence the vehicle responses at both DOFs are collected and fed into the output-only SID toolbox. The need to record the vehicle response at two locations simultaneously is a potential drawback. For example,  in a crowdsensing scenario, it may be impractical  to retrieve data at the tire level (unsprung mass) of the vehicle. Moreover, as mentioned in the former section, common vehicle suspension systems can have similar modal properties, specifically natural frequencies and mode shapes. The mechanical characteristics of various vehicles are shown in Table \ref{vcl_props}, and their tabulated modal responses and their average are presented in Table \ref{vcl_mdl_avg}. The properties shown in Table \ref{vcl_props} are selected from reference vehicles commonly used in the literature, which covers a wide range of linear vehicles \cite{sun2001modeling, bogsjo2012models, florin2013passive, gillespie1985measuring}. The fundamental frequency range of the vehicles also covers from 0.54 Hz to 36.78 Hz as shown in Table \ref{vcl_mdl_avg}, which is considerable, suggesting that these mode shapes are a good representation of available automobiles. As a simple demonstration, the average mode shapes are set as the identified values in the FRF. According to Equation \ref{eq:FRF3}, natural frequencies are also needed to produce a vehicle TF, as well as the mode shapes. The frequencies can be identified from the peaks of the PSD of the vehicle response at either DOF. In this section, bridge identification is performed using averaged mode shapes for the common vehicle by using FRF method.\par

\begin{table}[!htbp]
\centering
\caption{Mechanical properties of various vehicles}
\label{vcl_props}
\begin{tabular}{l|c|c|c|c|c|c}
\cline{2-6}
\multicolumn{1}{c|}{}                      & \multicolumn{5}{c|}{Vehicle ID}                                     & \textbf{}                           \\ \cline{2-7} 
\multicolumn{1}{c|}{}                      & \textbf{v1} & \textbf{v2} & \textbf{v3} & \textbf{v4} & \textbf{v5} & \multicolumn{1}{c|}{\textbf{Units}} \\ \hline
\multicolumn{1}{|l|}{Suspension Stiffness} & 1.8e6       & 62.30       & 128.7       & 2.7e5       & 5700        & \multicolumn{1}{c|}{N/m}            \\ \hline
\multicolumn{1}{|l|}{Suspension Damping}   & 1400        & 6.0         & 3.86        & 6000        & 290         & \multicolumn{1}{c|}{Ns/m}           \\ \hline
\multicolumn{1}{|l|}{Sprung Mass}          & 466.5       & 1.0         & 1.0         & 3400        & 466.5       & \multicolumn{1}{c|}{Kg}             \\ \hline
\multicolumn{1}{|l|}{Unsprung Mass}        & 49.8        & 0.15        & 0.162       & 350         & 49.8        & \multicolumn{1}{c|}{Kg}             \\ \hline
\multicolumn{1}{|l|}{Tire Stiffness}       & 7.2e5       & 653         & 643         & 9.5e5       & 1.35e5      & \multicolumn{1}{c|}{N/m}            \\ \hline
\multicolumn{1}{|l|}{Tire Damping}         & 0           & 0           & 0           & 300         & 1400        & \multicolumn{1}{c|}{Ns/m}           \\ \hline
\end{tabular}
\end{table}

\begin{figure}[!htbp]
\centering\includegraphics[width=0.9\linewidth]{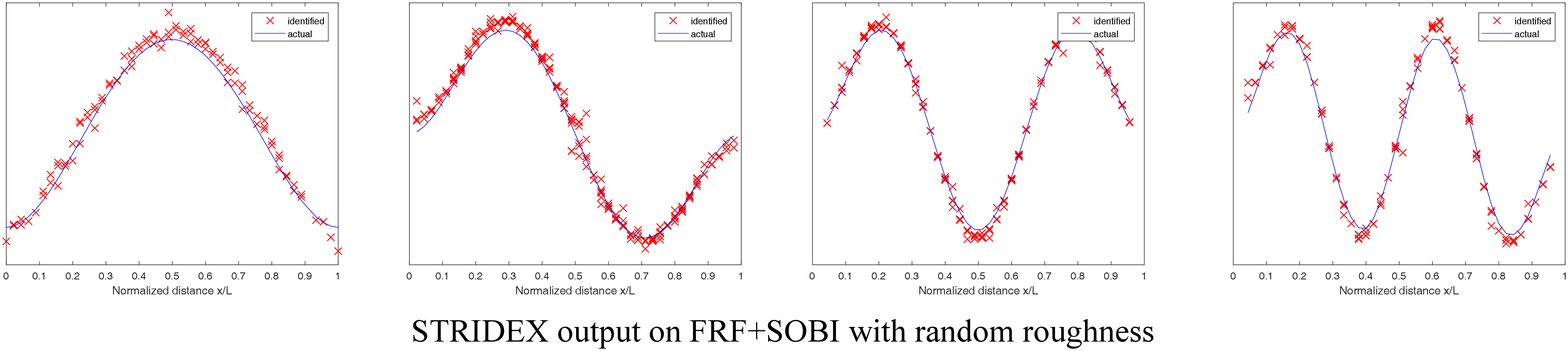}
\caption{Modal identification results - approximated common car}
\label{approx_rand}    
\end{figure}

In addition to the natural frequencies and the mode shapes, Equation \ref{eq:FRF3} requires damping ratio entries. In case of linear vehicles, like frequencies, these values are also identifiable by output-only SID of the system using only one channel vibration (sprung mass accelerations). A comparison between results shown in Figure \ref{approx_rand} with Figure \ref{stridex_2} confirm that the approximated transfer function is as successful as the actual one for the identification purposes. Identified frequencies shown in Table \ref{ident_frq_approx} show a very good match between estimated frequencies and actual ones. Estimated dampings presented in Table \ref{ident_dmp_approx} are acceptably accurate for modes $2$ to $4$, while the first mode still is far large. A comparison between Tables \ref{iden_dmp_1Hz} and \ref{ident_dmp_approx} show that the approximation for the mode shape slightly impacted negatively on the damping estimations. MAC values of the identified modes are also introduced in Table \ref{ident_mac_approx}. The range of MAC values are as desirable as the former case, promising that the approximation is generally successful.\par
These results show that using an approximated mode shape, the vehicle response can be characterized sufficiently using only one sensor in the vehicle cabin. Of course, it is preferable to utilize the mechanical properties provided by the manufacturer whenever possible.

\begin{table}[!htbp]
\centering
\caption{Vehicle modal characteristics}
\label{vcl_mdl_avg}
\begin{tabular}{c|c|c|c|c|c|c|}
\cline{2-7}
                                       & \multicolumn{2}{c|}{1st mode} & \multicolumn{2}{c|}{2nd mode}  & \multicolumn{2}{c|}{\begin{tabular}[c]{@{}c@{}}Natural\\ Frequencies\end{tabular}} \\ \cline{2-7} 
                                       & \textbf{DOF1} & \textbf{DOF2} & \textbf{DOF1}  & \textbf{DOF2} & \textbf{f1}                             & \textbf{f2}                              \\ \hline
\multicolumn{1}{|c|}{v1}               & 0.73          & 1.00          & -1.00          & 0.08          & 5.14                                    & 36.78                                    \\ \hline
\multicolumn{1}{|c|}{v2}               & 0.09          & 1.00          & -1.00          & 0.01          & 1.20                                    & 11.00                                    \\ \hline
\multicolumn{1}{|c|}{v3}               & 0.17          & 1.00          & -1.00          & 0.03          & 1.64                                    & 11.01                                    \\ \hline
\multicolumn{1}{|c|}{v4}               & 0.23          & 1.00          & -1.00          & 0.02          & 1.25                                    & 9.42                                     \\ \hline
\multicolumn{1}{|c|}{v5}               & 0.04          & 1.00          & -1.00          & 0.00          & 0.54                                    & 8.46                                     \\ \hline
\multicolumn{1}{|c|}{\textbf{Average}} & \textbf{0.25} & \textbf{1.00} & \textbf{-1.00} & \textbf{0.03} & \textbf{1.96}                           & \textbf{15.33}                           \\ \hline
\end{tabular}
\end{table}

\begin{table}[!htbp]
\centering
\caption{Identified frequencies using approximated common car}
\label{ident_frq_approx}
\begin{tabular}{cl|c|c|c|}
\cline{3-5}
\multicolumn{2}{c}{\textit{values in Hz}}          & \multicolumn{3}{|c|}{FRF+SOBI} \\ \hline
\multicolumn{1}{|c|}{Mode ID}    & \textbf{Actual} & Sinusoidal  & EJ     & Random \\ \hline
\multicolumn{1}{|c|}{\textbf{1}} & \textbf{0.1357} & 0.1324      & 0.1421 & 0.1327 \\ \hline
\multicolumn{1}{|c|}{\textbf{2}} & \textbf{0.3714} & 0.3687      & 0.3686 & 0.3702 \\ \hline
\multicolumn{1}{|c|}{\textbf{3}} & \textbf{0.7213} & 0.7211      & 0.7215 & 0.7196 \\ \hline
\multicolumn{1}{|c|}{\textbf{4}} & \textbf{1.1710} & 1.1861      & 1.1839 & 1.1793 \\ \hline
\end{tabular}
\end{table}

\begin{table}[!htbp]
\centering
\caption{Identified damping ratios using approximated common car}
\label{ident_dmp_approx}
\begin{tabular}{cc|c|c|c|}
\cline{3-5}
\multicolumn{2}{c}{\textit{values in \%}}                               & \multicolumn{3}{|c|}{FRF+SOBI} \\ \hline
\multicolumn{1}{|c|}{Mode ID}    & \multicolumn{1}{l|}{\textbf{Actual}} & Sinusoidal  & EJ     & Random \\ \hline
\multicolumn{1}{|c|}{\textbf{1}} & \textbf{2.00}                        & 21.79       & 26.07  & 14.18  \\ \hline
\multicolumn{1}{|c|}{\textbf{2}} & \textbf{0.80}                        & 1.75        & 1.70   & 1.67   \\ \hline
\multicolumn{1}{|c|}{\textbf{3}} & \textbf{0.60}                        & 1.59        & 0.53   & 0.47   \\ \hline
\multicolumn{1}{|c|}{\textbf{4}} & \textbf{0.70}                        & 1.04        & 0.79   & 0.35   \\ \hline
\end{tabular}
\end{table}

\begin{table}[!htbp]
\centering
\caption{Identified modal assurance criteria (MAC) using approximated common car}
\label{ident_mac_approx}
\begin{tabular}{c|c|c|c|}
\cline{2-4}
\textit{values in \%}            & \multicolumn{3}{c|}{FRF+SOBI} \\ \hline
\multicolumn{1}{|c|}{Mode ID}    & Sinusoidal  & EJ     & Random \\ \hline
\multicolumn{1}{|c|}{\textbf{1}} & 0.9915       & 0.9888  & 0.9966  \\ \hline
\multicolumn{1}{|c|}{\textbf{2}} & 0.9922       & 0.9904  & 0.9893  \\ \hline
\multicolumn{1}{|c|}{\textbf{3}} & 0.9806       & 0.9737  & 0.9803  \\ \hline
\multicolumn{1}{|c|}{\textbf{4}} & 0.9722       & 0.9723  & 0.9761  \\ \hline
\end{tabular}
\end{table}

\section{Conclusion}

In this study, two methods were proposed for the comprehensive bridge system identification using vehicle-carried sensor data. In the first approach, vehicle deconvolution using the vehicle frequency response function (FRF), along with second order blind identification (SOBI) extracted bridge vibration from mixed signals collected by drive-by vehicles. Empirical modal decomposition was proposed as an alternative approach for vehicle deconvolution. Throughout the extraction phase, vehicle suspension effects and roughness-induced vibrations were removed. Finally, for bridge system identification (SID), resulting signal (from either the FRF or the EEMD method) represents pure mobile sensing data and was processed by the extended structural identification using expectation maximization (STRIDEX) algorithm for bridge modal identification. Numerical case studies from a $500 m$ long bridge were used to validate proposed methods. \par

The methods were both successful in estimating first four modes of the bridge. Modal assurance criteria (MAC) values for the estimated mode shapes from both methods were all above 0.94. In terms of the estimated frequencies, estimated values of FRF and EEMD methods had all less than $1.2\%$ and $1.3\%$ error from the actual values, respectively. The accuracy of the damping ratios was generally on par with traditional SID methods; in some cases, the estimates were near exact, e.g., the second mode. \par
In order to investigate methods' robustness to the sensing vehicle properties, a second property set (properties of common vehicles) was evaluated. Both identification methods were applied on the measured data from this vehicle. In this case, modal identification results were not considerably affected with respect to the case of the designated vehicle property. However, estimated damping ratios of the first mode were affected significantly. The possible reasons were discussed.\par

The EEMD-based method operates without vehicle property information. Overall, the FRF approach yielded more accurate SID results of the bridge. As a means to circumvent the vehicle SID phase in the FRF method, a simplified procedure was proposed to approximate vehicle transfer function using data exclusively collected from within the cabin. The results showed that this method was successful in producing accurate modal property estimate; the frequency error was at max $1.2\%$ and the MAC values were above 0.97.\par

The rate at which the SHM community can incorporate information extracted from crowdsourced data depends on how its analytical tools can adopt to new, more readily available data formats, e.g., mobile sensing data. The proposed methods enable robust extraction of important bridge information using data types that are compatible with large-scale vehicles networks. An ability to turn everyday vehicle-based data sets into core SHM information fuels more regular observations on the operational behavior of the bridge, which in turn supports more frequent condition reports. 

\section{Acknowledgments}

Research funding is partially provided by the National Science Foundation through Grant No. CMMI-1351537 by the Hazard Mitigation and Structural Engineering program, and by a grant from the Commonwealth of Pennsylvania, Department of Community and Economic Development, through the Pennsylvania Infrastructure Technology Alliance (PITA). We would also like to thank Eng. Araz Sadoughi Shabestari, bridge designer, for his feedback.


\section{Appendix A: Discussion on the Generality of the Methods}\label{append0:short_bridge}
In this part, we implement the proposed pipelines on a shorter bridge ($300$m single span) with a road roughness profile adopted from ISO8608:2016 \cite{iso20168608} (considering road class C). In this case, the possibility of the methods is examined in (a) a shorter bridge, (b) with a realistic road profile roughness spectrum, and (c) with closely-spaced vehicle and the bridge frequency contents. The bridge is supported by rigid constraints from both ends and is modeled as a $5,000$ DOF concrete beam in Opensees. The nodal mass of each DOFs is set to $1039$ Kg with $10.4m^2$ and $51.5m^4$ cross-sectional area and moment of inertia, respectively. Based on these mechanical properties, the first four natural modes of the bridge have the following frequencies: $0.40$ Hz, $1.11$ Hz, $2.18$ Hz, and $3.61$ Hz. The vehicle properties are adopted based on Table \ref{table:1} and the sensing pattern is identical with the previous cases. From the frequency values, the vehicle natural mode is located closely after the fourth natural mode of the bridge which adds complexity, especially for applying Method 2. 

\begin{figure}[!ht]
\centering\includegraphics[width=0.5\linewidth]{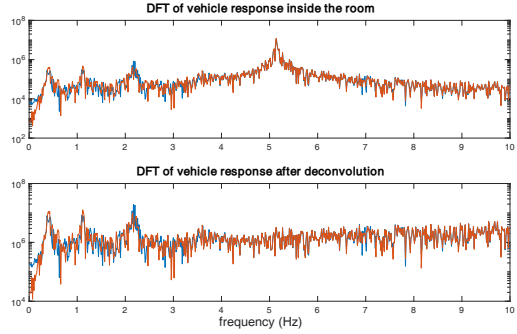}
\caption{FRF deconvolution results for ISO8606 roughness and 300m bridge}
\label{ISO300_deconv}
\end{figure}

\begin{figure}[!htbp]
\centering\includegraphics[width=0.9\linewidth]{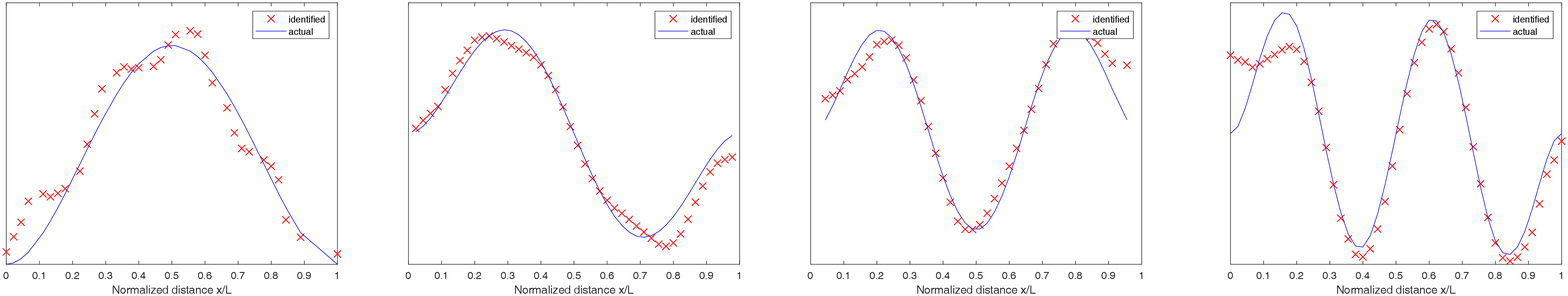}
\caption{FRF+SOBI with ISO8606 roughness and 300m bridge}
\label{ISO300_Method1_modes}    
\end{figure}

Figure \ref{ISO300_deconv} shows the deconvolution effect on the vehicle response FRF and Figure \ref{ISO300_Method1_modes} presents identification results from applying Method 1. Deconvolution using the transfer function could extract the highly-damped mode at frequency about $5$Hz that associates with the vehicle fundamental model. In addition, from Figure \ref{ISO300_Method1_modes}, the natural mode shapes are acceptably identified (MAC values ranging from $0.9195$ to $0.9718$). \par
\begin{figure}[!ht]
\centering\includegraphics[width=0.5\linewidth]{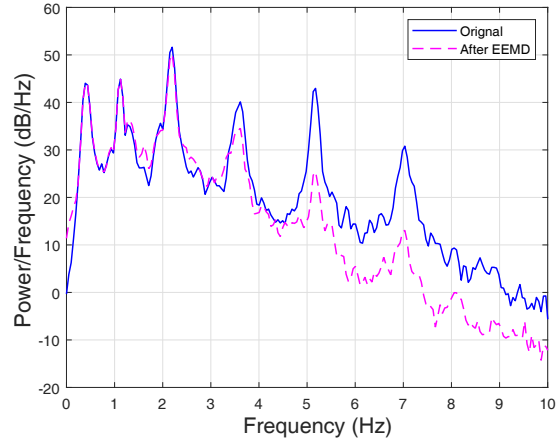}
\caption{EEMD deconvolution results for ISO8606 roughness and 300m bridge}
\label{ISO300_deconvEEMD}
\end{figure}
\begin{figure}[!htbp]
\centering\includegraphics[width=0.9\linewidth]{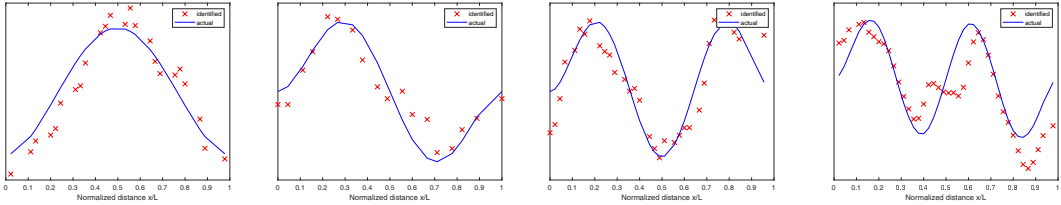}
\caption{EEMD+SOBI with ISO8606 roughness and 300m bridge}
\label{ISO300_Method2_modes}    
\end{figure}

For Method 2, the primary challenge is that the $4^{th}$ natural mode of the bridge is very close to the frequency content of the vehicle and EEMD is not able to perfectly extract the vehicle mode out the mixed signal. 
Modal identification results from Method 2 are presented in Figures \ref{ISO300_deconvEEMD} and \ref{ISO300_Method2_modes}. Figure \ref{ISO300_deconvEEMD} demonstrated that EEMD-based method can be successful for the deconvolution tasks in cases with closely-spaced frequencies. Identification results also are fairly acceptable with MAC values ranging from $0.7870$ for the fourth mode to $0.9185$ for the first mode (as expected, mode 4 has the least accuracy because of its closeness to the vehicle frequency). Generally it is concluded that the proposed methods are suitable for a wide variety of bridges, however, the frequency contents proximity or overlap has a detrimental effect on the final output.  

\section{Appendix B: Modal Aggregation} \label{append1:mode_aggregate}
In STRIDEX, MSR function maps mobile measurements to stationary virtual probing locations (VPLs) and iteratively calculate the best estimate for structural parameters using expectation maximization (EM) algorithm. In the original studies of STRIDEX, the minimum model size is selected such that the number of VPLs is equal to the number of moving sensors. The number of VPLs can be set to $p\times N$, where $N$ is the number of mobile sensors and $p$ is the model order. With an increased $p$, a higher spatial resolution is produced for the identified mode shapes can be achieved. For this study, however, the same number of VPLs as mobile sensors has been considered. Four model orders, $p = 1,2,3$, and $4$ have been analyzed here. For each model order, 11 VPL sets have been presumed and superposed to enhance mode shape resolutions. Each VPL set consists of eight VPLs with equal spacing, and with a certain eccentricity. In total, for each analysis, $11 \times 4=44$ output sets exist, each consists of estimated modal properties. For aggregating these, an automated algorithm has been adopted. For a certain desired mode, the algorithm automatically produces a perfect sine-shaped mode shape estimation (e.g., half-sine for mode one, full-sine for mode two), and check the dissimilarity between identified mode shape and this preliminary estimation for each file. Those sets that have dissimilarity measures less than a certain threshold (used 0.72 to 0.74), are selected and combined to shape the final identified mode shapes. The dissimilarity measure used in this study is the second norm of $y - \tilde{y}$ or $y + \tilde{y}$. If the second expression governs, it means that the mode shape is accurate but inverted. In addition to the dissimilarity condition for output selection, sets with damping ratios over $\%30$ are excluded as well. This automated algorithm accelerate mode shape reconstruction process significantly and keep it more controlled.

\section{Appendix C: Auxiliary Identifications - Road Profile and Vehicle Suspension} \label{append2:auxiliary}

\begin{figure}[!ht]
\centering\includegraphics[width=1.0\linewidth]{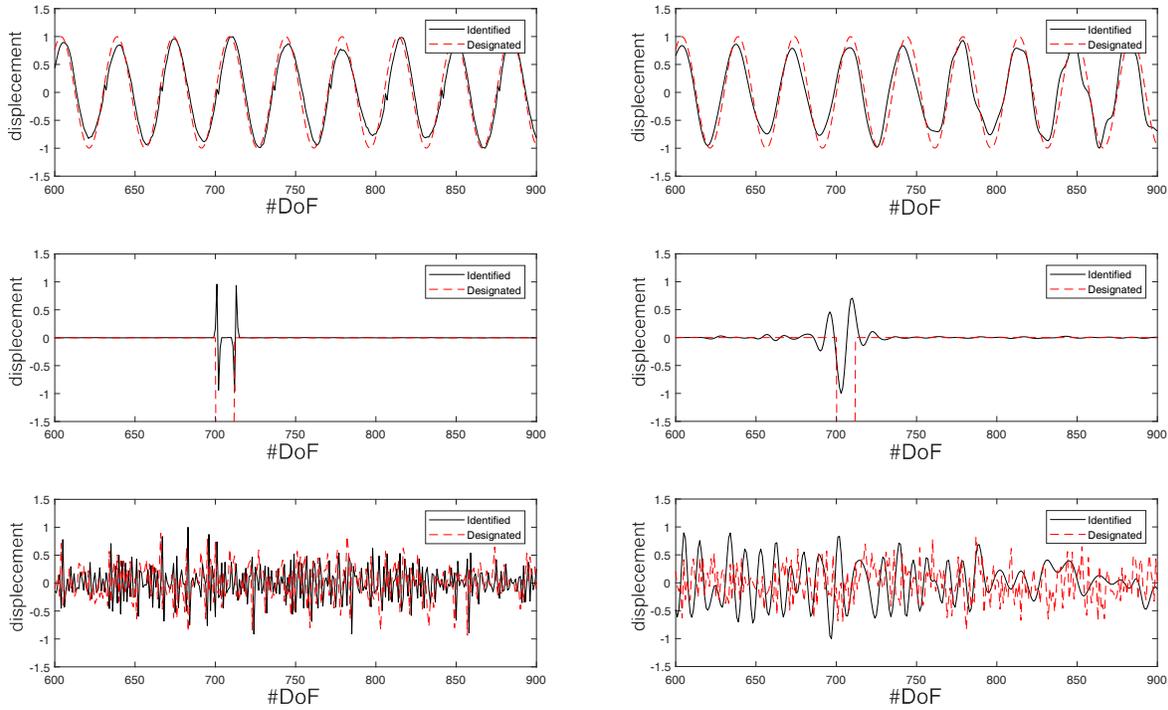}
\caption{Roughness profile estimation - FRF method on left, EEMD on Right}
\label{rgh_ident}
\end{figure}

The procedure of bridge signal extraction consists of distinct steps for vehicle effect separation and roughness-caused vibration separation. With this notion, one may be interested to exploit byproducts of the processes for different objectives, i.e. road roughness profile estimation and vehicle suspension identification. SOBI as the linear mix separator of both methods computes two signal channels, one of which was used as bridge vibration for SID. The other channel theoretically shall correspond to the road roughness profile. In Figure \ref{rgh_ident} the second extracted channels from SOBI (identified roughness profiles) are compared with three predesignated profiles. 

\begin{figure}[!htbp]
\centering\includegraphics[width=1.0\linewidth]{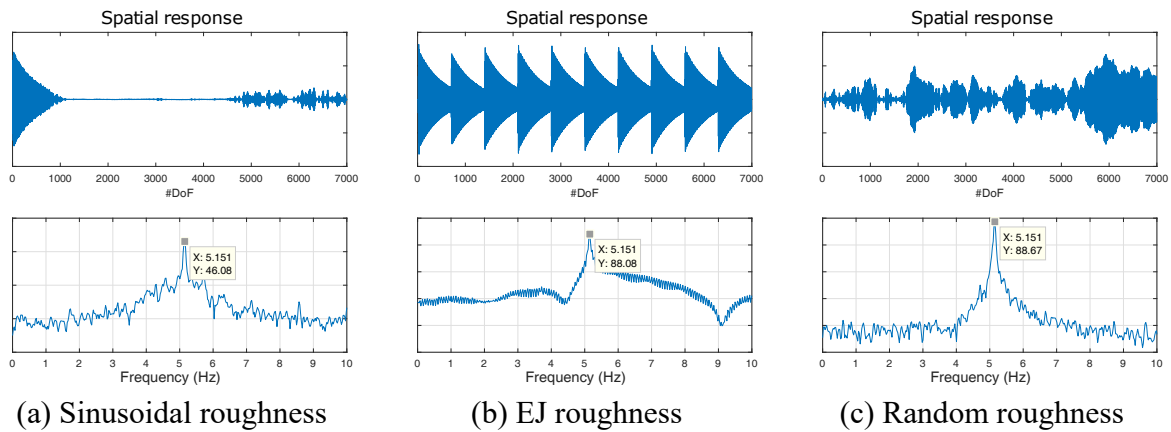}
\caption{IMF corresponding to vehicle - spatial and frequency representation}
\label{vcl_ident_IMF}
\end{figure}

Location domain shown in Figure \ref{rgh_ident} is just a small portion of the entire path that is selected for more clarity. FRF approach is more successful in accurately estimating roughness profile. In fact, in the case of random profile, EEMD method does not extract an acceptable signal as the road profile estimation. An important difference between the two approaches in this study is the necessity of pre-identification of the vehicle. In fact, FRF approach needs vehicle characteristics as a given which was created by a preliminary SID of the vehicle while traveling over normal roads. However, in EEMD approach, vehicle identification is not required and the method can produce the vehicle effect as an IMF. Therefore, this IMF is assumed to be the pure vehicle response and by representing it into the frequency domain, the vehicle natural frequencies shall be identified. Figure \ref{vcl_ident_IMF} shows these analyses for the three cases of interest. 

Figure \ref{vcl_ident_IMF} illustrates that the IMF extracted as the vehicle response is exactly showing a large spike on the location of the vehicle fundamental mode. Note that because of the low sampling frequency for the discrete simulation of the vehicle, the second natural frequency is not clearly visible.


\bibliography{manuscript}







\section*{List of Symbols}
\addcontentsline{toc}{chapter}{List of Symbols}

\begin{tabular}{cp{0.8\textwidth}}
  $x_k$ & state vector at time step k \\
  $A$ & state matrix \\
  $\eta_k$ & systematic noise at time step k (bridge random load) \\
  $y_k$ & observation vector at time step k \\
  $C$ & observation matrix \\
  $\nu_k$ & sensing noise at time step k \\
  $Q$ & covariance matrix of the systematic noise \\
  $R$ & covariance matrix of the sensing noise \\
  $p$ & model order \\
  $N_\alpha$ & number of virtual probing nodes \\
  $N_0$ & number of observation nodes \\
  $\Omega_k$ & mode shape regression function at time step k \\
  $s_k^0$ & position vector of sensing nodes at time step k \\
  $s_i^\alpha$ & location of the \textit{i}th virtual probing location \\
  $\Delta s^\alpha$ & uniform distance between virtual probing locations \\
  $y_k$, $y_k^{br}$ & pure bridge dynamic response at time step k \\
  $y_k^{act}$ & actual response recieved by the sensor at time step k \\
  $y_k^{if}$ & bridge response to the vehicle-bridge interaction force at time step k \\
  $y_k^{vbi}$ & actual bridge response including vehicle-bridge interaction at time step k \\
  $y_k^{act}$ & actual response recieved by the sensor at time step k \\
  $y_k^{rgh}$ & road profile roughness displacement under the tire at time step k \\
  $y_k^{eng}$ & engine-induced vibration at time step k \\
  $y_k^{obs}$ & observed vibration within the vehicle at time step k \\
  $y_k^{inp}$ & displacement input of the vehicle at the tire level at time step k \\
  $Y(\omega)$ & frequency representation of $y_k$ \\
  $H(\omega)$ & vehicle transfer function \\
  $\alpha(\omega)$ & frequency response function \\
  $\Phi$ & vehicle mode shapes \\
  $m_s$ & vehicle sprung mass \\
  $m_{us}$ & vehicle unsprung mass \\
  $c_s$ & vehicle sprung damping \\
  $c_{us}$ & vehicle unsprung damping \\
  $k_s$ & vehicle suspension stifness \\
  $k_{us}$ & vehicle tire stiffness \\
\end{tabular}\\

\end{document}